\documentclass[11pt,a4paper]{article}

\usepackage[text={176mm,240mm},centering]{geometry}
\usepackage{graphicx,color,float,amsmath,amssymb,amsfonts,cite,multirow,multicol,subcaption,ulem,bbm,physics,amstext,mathtools,nicematrix,authblk}
\usepackage[colorlinks=true,linkcolor=red,breaklinks=true,urlcolor=blue,citecolor=blue]{hyperref}
\usepackage{genyoungtabtikz}

\newcolumntype{C}{>{$}c<{$}}
\newcolumntype{L}{>{$}l<{$}}
\newcolumntype{R}{>{$}r<{$}}

\newcommand{\doublet}[2]{\arraycolsep1pt\left(\begin{array}{c} #1\\ #2\\ \end{array}\right)}
\newcommand{\matrixTwo}[4]{\arraycolsep1pt\left(\begin{array}{cc} #1 & #2\\ #3 & #4\\ \end{array}\right)}
\newcommand{\triplet}[3]{\arraycolsep1pt\left(\begin{array}{c} #1\\ #2\\ #3\\ \end{array}\right)}
\newcommand{\Range}[2]{#1_1,#1_2,\ldots,#1_{#2}}

\Yboxdim{12pt}

\begin{document}

\title{Phase Conventions in Hadron Physics from the Perspective of the Quark Model}

\date{\today}




\author{
     Yu Lu$^{1}$
        \footnote{Email address:\texttt{ylu@ucas.ac.cn} }~ ,
     Hao-Jie Jing$^{2}$
        \footnote{Email address:\texttt{jinghaojie@sxu.edu.cn} }~ ,
     Jia-Jun Wu$^{1,3}$
        \footnote{Email address:\texttt{wujiajun@ucas.ac.cn}, corresponding author} 
        \thanks{Yu Lu and Hao-Jie Jing contributed equally to this work.}\\
      {\it\small$^1$ School of Physical Sciences, University of Chinese Academy of Sciences (UCAS), Beijing 100049, China}\\
      {\it\small$^2$ College of Physics and Electronic Engineering, Shanxi University, Taiyuan 030006, China}\\
      {\it\small$^3$ Southern Center for Nuclear-Science Theory (SCNT), Institute of Modern Physics, Chinese Academy of Sciences, Huizhou 516000, China}\\
}

\maketitle

\abstract{
Convenient and consistent phase convention is important in the construction of the hadronic Lagrangian.
However, the importance of phase convention has been overlooked for a long time and the sources of different conventions are never explicitly addressed.
This obscure situation can cause mistakes and misinterpretations in hadron physics.
In this paper, we systematically analyze and compare the flavor $SU_3$ phase conventions from the perspective of the quark model. 
All sources which could lead to different conventions are pointed out and carefully studied.
With the tool of the quark model, we also clarify some misconceptions and 
demonstrate a consistent way to incorporate different conventions.}


\thispagestyle{empty}

\section{Introduction}

Quantum mechanics is built upon the Hilbert space, where two vectors $\psi,\phi$ can be linearly combined into a new state.
$a\psi_1 + b\psi_2$ is generally not the same as $a\psi_1 - b\psi_2$.
The change of the sign at the amplitude level results in different interference term which leads to different physical predictions.
Thus, every physicist agrees that the relative phase between the two vectors are important.
On the other hand, the overall phase, such as the complex $\eta$ in $\eta(a\psi_1 + b\psi_2)$ can be set arbitrarily because it is not physically observable.
Despite this degree of freedom in setting the arbitrary overall phases, a unified convention will undoubtedly be helpful, 
especially when comparing results from various sources.

For simpler groups, such as the $SU_2$ group of the angular momentum, there is a widely accepted phase convention for physicists, the renowned Condon-Shortley phase convention.
For larger groups, despite the existing of the natural extension of the Condon-Shortley phase convention in mathematics~\cite{Biedenharn:1963zk, Baird:1963wv, Baird:1964zm, Baird:1964zn}, different physicist starts to invent and stick to their own phase conventions. 
%

In principle, it is correct that all phase conventions are physically equivalent as long as each convention is self-consistent,
and some peculiar conventions should be suitably explained once used.
However, there are inevitably temporary treatments which make the conventions hard to track, e.g., it may happen that not all the $SU_3$ multiplets are of interest, and only some $SU_2$ slices of the full $SU_3$ multiplets are calculated for physical convenience.
What appears to be the irrelevant overall factors for $SU_2$ are in fact deeply connected by $SU_3$, thus, they are essentially the crucial relative phase.

Differences in conventions and the temporary treatment mentioned above have made it practically challenging to compare and merge coupling constants from different sources.
This situation also greatly hinders the communication of physicists.
In practice, it is prone to introduce inconsistencies to the convention, however, those inconsistencies can sometimes be absorbed by the redefinition of hadronic fields or the coupling constants in the Lagrangian.
This brings additional complexity in checking and comparing the results among literature.
This chaotic situation was pointed out and a recommended convention is offered in Ref.~\cite{Chen:1979qz}, however, the detailed analysis and comparison of different sources is still missing.
It would be beneficial if the intricate conventions can be classified or compared, and different origins of the conventions can be addressed systematically.

This is the topic that this paper mainly devotes to.
To facilitate the analysis, we use the quark model which is familiar by physicists as a proxy to the group theory.
With the quark model, we will address the various conventions which occur at different levels and stages and offer a systematic way to pinpoint and compare the intricate conventions.
We will show that the convention is not just from mathematics, while it is a result of interplay between mathematics and physics.
In this paper, we mainly focus on the $SU_3$ group in the hadron flavor degree of freedom with a slight extension to $SU_4$.
We also show a interesting result coming from the constrain of the color degree of freedom.

\begin{figure}[!htbp]
    \centering
    \includegraphics[width=0.6\textwidth]{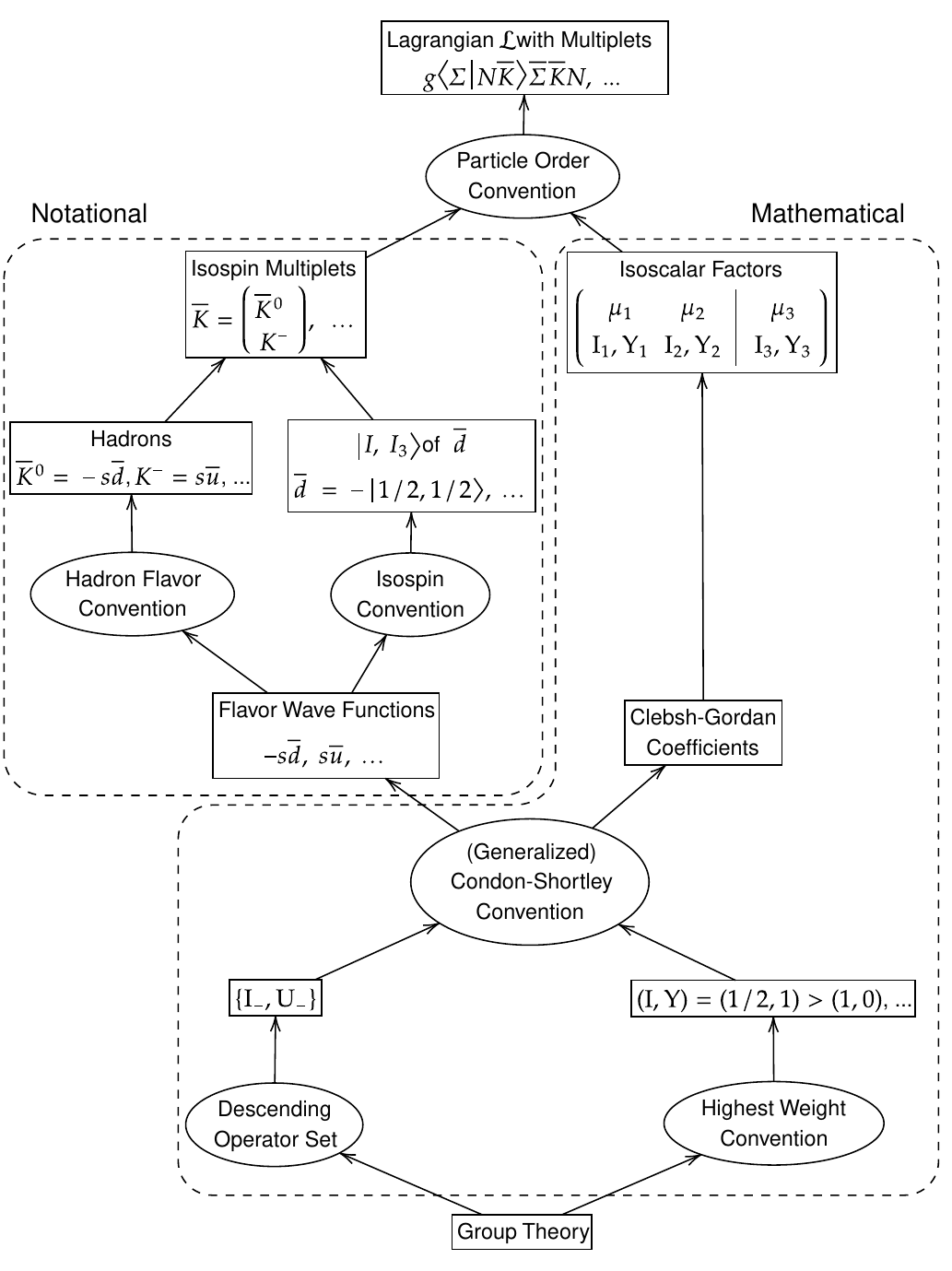}
    \caption{The workflow of writing down the Lagrangian, where the ellipses mark the conventions which leads to different rectangles outcomes.}
    \label{fig:flowchart}
\end{figure}

We summarize the whole procedure to write down hadronic Lagrangian in the Figure~\ref{fig:flowchart}.
The whole theme starting from the flavor wave functions part is the identification of hadrons with derived wave functions.
In this part, the differences between different conventions are pure notational.
In principle, it is not difficult to translate different conventions by the redefinition of the hadronic field.
Despite of its superficial nature, it may lead to various confusions and misinterpretations.

This paper is organized as the following.
Section~\ref{sec:cgcs} is devoted to the group theory, where different generalizations of the Condon-Shortley phase conventions of the Clebsh-Gordan coefficients are discussed.
We also show an interesting result from the interplay of flavor and color degree of freedom.
Section~\ref{sec:quark} explains the group theory result with the language of the quark model and different Hadron flavor conventions are derived and compared.
The isoscalar factor under the convention of Chen et.al.\cite{Chen:2002gd} is derived in Section~\ref{sec:isf}, which is also compared with the convention used by de Swart \cite{deSwart:1963pdg, PDG} and Rabl et.al.\cite{Rabl:1975zy, Haacke:1975rt}.
We give a short summary of this paper in Section~\ref{sec:summary} and some calculation details in Appendix~\ref{appendix}, \ref{sec:Decuplet_matrix}.


\section{Clebsh-Gordan Coefficients}\label{sec:cgcs}

\subsection{$SU_2$ and Condon-Shortley Phase Convention}

\begin{figure}[!htb]
\centering
\tikzset{every picture/.style={line width=0.6pt}} 

\begin{tikzpicture}[x=0.75pt,y=0.75pt,yscale=-0.6,xscale=0.6, every node/.style={scale=0.6}]

\draw (158,42.5) node    {$\ket{J_{1} +J_{2} ,J_{1} +J_{2}}$};
\draw (158.5,126.5) node    {$\ket{J_{1} +J_{2} ,J_{1} +J_{2} -1}$};
\draw (160,393.5) node    {$\ket{J_{1} +J_{2} ,-J_{1} -J_{2}}$};
\draw (412.5,126.5) node    {$\ket{J_{1} +J_{2} -1,J_{1} +J_{2} -1}$};
\draw (568,182) node    {$\dotsc $};
\draw (159,230) node    {$\dotsc $};
\draw (159.5,322.5) node    {$\ket{J_{1} +J_{2} ,-J_{1} -J_{2} +1}$};
\draw (413.5,180.5) node    {$\ket{J_{1} +J_{2} -1,J_{1} +J_{2} -2}$};
\draw (168,70.4) node [anchor=north west][inner sep=0.75pt]    {$J_{-}$};
\draw (240,99) node [anchor=north west][inner sep=0.75pt]   [align=left] {Orthogonal};
\draw (414,323.5) node    {$\ket{J_{1} +J_{2} -1,-J_{1} -J_{2} +1}$};
\draw (414,231) node    {$\dotsc $};
\draw (568,247) node    {$\dotsc $};
\draw    (158.08,56) -- (158.41,111) ;
\draw [shift={(158.42,113)}, rotate = 269.66] [color={rgb, 255:red, 0; green, 0; blue, 0 }  ][line width=0.75]    (10.93,-3.29) .. controls (6.95,-1.4) and (3.31,-0.3) .. (0,0) .. controls (3.31,0.3) and (6.95,1.4) .. (10.93,3.29)   ;
\draw  [dash pattern={on 4.5pt off 4.5pt}]  (158.57,140) -- (158.93,216) ;
\draw [shift={(158.94,218)}, rotate = 269.72] [color={rgb, 255:red, 0; green, 0; blue, 0 }  ][line width=0.75]    (10.93,-3.29) .. controls (6.95,-1.4) and (3.31,-0.3) .. (0,0) .. controls (3.31,0.3) and (6.95,1.4) .. (10.93,3.29)   ;
\draw  [dash pattern={on 4.5pt off 4.5pt}]  (159.6,336) -- (159.89,378) ;
\draw [shift={(159.9,380)}, rotate = 269.6] [color={rgb, 255:red, 0; green, 0; blue, 0 }  ][line width=0.75]    (10.93,-3.29) .. controls (6.95,-1.4) and (3.31,-0.3) .. (0,0) .. controls (3.31,0.3) and (6.95,1.4) .. (10.93,3.29)   ;
\draw    (159.06,242) -- (159.42,307) ;
\draw [shift={(159.43,309)}, rotate = 269.69] [color={rgb, 255:red, 0; green, 0; blue, 0 }  ][line width=0.75]    (10.93,-3.29) .. controls (6.95,-1.4) and (3.31,-0.3) .. (0,0) .. controls (3.31,0.3) and (6.95,1.4) .. (10.93,3.29)   ;
\draw    (231,125) -- (319,125)(231,128) -- (319,128) ;
\draw [shift={(327,126.5)}, rotate = 180] [color={rgb, 255:red, 0; green, 0; blue, 0 }  ][line width=0.75]    (10.93,-3.29) .. controls (6.95,-1.4) and (3.31,-0.3) .. (0,0) .. controls (3.31,0.3) and (6.95,1.4) .. (10.93,3.29)   ;
\draw    (412.75,140) -- (413.21,165) ;
\draw [shift={(413.25,167)}, rotate = 268.94] [color={rgb, 255:red, 0; green, 0; blue, 0 }  ][line width=0.75]    (10.93,-3.29) .. controls (6.95,-1.4) and (3.31,-0.3) .. (0,0) .. controls (3.31,0.3) and (6.95,1.4) .. (10.93,3.29)   ;
\draw    (499.01,179.83) -- (544.01,180.27)(498.99,182.83) -- (543.99,183.27) ;
\draw [shift={(552,181.84)}, rotate = 180.56] [color={rgb, 255:red, 0; green, 0; blue, 0 }  ][line width=0.75]    (10.93,-3.29) .. controls (6.95,-1.4) and (3.31,-0.3) .. (0,0) .. controls (3.31,0.3) and (6.95,1.4) .. (10.93,3.29)   ;
\draw    (413.63,194) -- (413.86,217) ;
\draw [shift={(413.88,219)}, rotate = 269.43] [color={rgb, 255:red, 0; green, 0; blue, 0 }  ][line width=0.75]    (10.93,-3.29) .. controls (6.95,-1.4) and (3.31,-0.3) .. (0,0) .. controls (3.31,0.3) and (6.95,1.4) .. (10.93,3.29)   ;
\draw    (414,243) -- (414,308) ;
\draw [shift={(414,310)}, rotate = 270] [color={rgb, 255:red, 0; green, 0; blue, 0 }  ][line width=0.75]    (10.93,-3.29) .. controls (6.95,-1.4) and (3.31,-0.3) .. (0,0) .. controls (3.31,0.3) and (6.95,1.4) .. (10.93,3.29)   ;
\draw    (568,194) -- (568,233) ;
\draw [shift={(568,235)}, rotate = 270] [color={rgb, 255:red, 0; green, 0; blue, 0 }  ][line width=0.75]    (10.93,-3.29) .. controls (6.95,-1.4) and (3.31,-0.3) .. (0,0) .. controls (3.31,0.3) and (6.95,1.4) .. (10.93,3.29)   ;

\end{tikzpicture} 
  \caption{Workflow of getting $SU_2$ Clebsch-Gordan coefficients by descending operator $J_{-}$.}
\label{fig:SU2}
\end{figure}
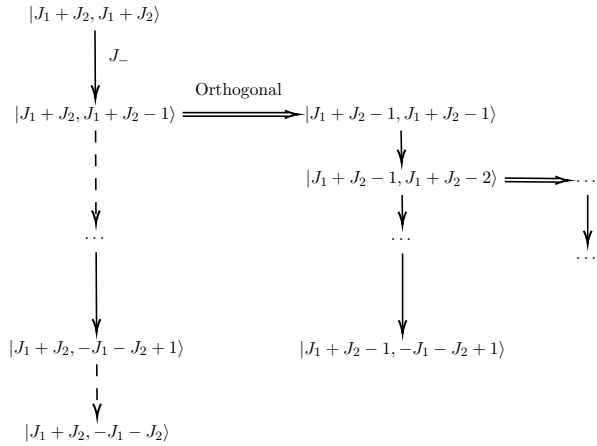

The traditional way to get the $SU_2$ Clebsh-Gordan coefficients (CGCs) is the descending operator method.
%
%
The main idea is a combination of descending operator $J_-$ with orthogonalization.
The details can be found in many quantum mechanics textbooks, such as Chapt. 3.7 in Ref.~\cite{Sakurai:1994}.
Since this method will be extended to $SU_3$, 
we demonstrate the key steps in the following.

The matrix element of $J_{\pm}$ operator is derived from the Casimir operator $\vec{J}^2$ of $SU_2$ which has diagonal matrix form.
The $J_{\pm}$ is constructed to be a Hermitian conjugated pair, and by the assumption that both matrix elements should be positive, one can take the square root of the diagonal and get the matrix element.
Specifically,
\begin{align}
    J_{+} J_{-} &= J^2 - J_z^2 + J_z, \\
    \bra{jm} J_+ J_- \ket{jm} &= j(j+1)- m(m-1),\\
    J_- \ket{jm} &= \sqrt{j(j+1) - m (m-1)} \ket{jm-1}.
\end{align}

Starting from the highest weight $\ket{J,J_z}=\ket{J_1 + J_2, J_1 +J_2}$, one recursively applies the $J_-$ operator which gradually decrease $J_z$ by one unit.
The matrix element of $J_-$ operator is conventionally assumed to be positive.
The process naturally terminates when reaching lowest weights $\ket{J,J_z}=\ket{J_1 + J_2, -J_1 -J_2}$.
All the signs before the family $\ket{J_1+J_2,m}, m=(-(J_1+J_2),..., J_1+J_2)$ are fixed by (in fact, the same as) the highest weight $\ket{J_1+J_2,J_1+J_2}$.
When this highest weight is expanded in the uncoupled basis, the expanding coefficient (CGC) is assumed to be $+1$, i.e. $\ket{J_1+J_2, J_1+J_2} = \ket{J_1,J_1}\ket{J_2,J_2}$.
In this uncoupled basis, the $J_-$ operator works as
\begin{align}
    J_- (\ket{J_1,m_1}\ket{J_2,m_2}) = (J_- \ket{J_1,m_1})\ket{J_2,m_2} + \ket{J_1,m_1}(J_- \ket{J_2,m_2}).
\end{align}
As a result, the CGCs within this $J=J_1+J_2$ family can be fixed.

To get the rest of the CGCs, one has to first make an assumption about the signs of $\ket{J_1+J_2-1, J_1+J_2-1}$, which is the highest weight of $J=J_1+J_2-1$ family.
Clearly, $\ket{J_1+J_2-1, J_1+J_2-1}$ should be orthogonal to $\ket{J_1+J_2, J_1+J_2-1}$, which will fix the CGCs up to an overall sign.
This sign can be fixed again by requiring the first non-zero CGC to be positive, i.e. $\braket{J_1,J_1,J_2,J_2 -1}{J_1+J_2-1,J_1+J_2-1}>0$.
And one again recursively applies the $J_-$ operator to $\ket{J_1+J_2-1, J_1+J_2-1}$, 
and the procedure goes on until all the CGCs are worked out.
We summarized this descending operator method in Figure~\ref{fig:SU2}.

The convention that 
$\braket{J_1,J_1,J_2, J-J_1 }{J,J}>0, J=(|J_1-J_2|, \ldots, J_1+J_2)$
is the renowned Condon-Shortley phase convention.

\subsection{Generalized Condon-Shortley Phase Convention for $SU_3$}\label{sec:su3cgc}

Theoretically, one can apply this descending operator method to get the CGCs for $SU_n$, which consists of the following two steps. 
\begin{enumerate}
    \item Selecting a complete set of descending operators, whose matrix elements are set to be positive.
    \item Extending the Condon-Shortley phase conventions in the orthogonalization process.
\end{enumerate}
There are various ways to achieve this, which leads to different conventions.
%

Like the gauges in quantum field theory, all conventions are mathematically equivalent, and they should lead to the same prediction on the physical observable.
Despite the equivalence of the conventions, it turns out that some choices are mathematically more elegant and more convenient to generalize.
In this work, we begin with the analysis of the first prerequisite, namely, the selection of the descending operators.

\begin{figure}[h]
  \centering
  \begin{subfigure}{0.3\textwidth}
    \centering
    \tikzset{every picture/.style={line width=0.75pt}} 

\begin{minipage}{\linewidth}
\centering

\scalebox{0.5}{

\begin{tikzpicture}[x=0.75pt,y=0.75pt,yscale=-1,xscale=1]


\draw (214,247.02) node [anchor=north west][inner sep=0.75pt]    {$V_{-}$};
\draw (181,187.02) node [anchor=north west][inner sep=0.75pt]    {$I_{-}$};
\draw (134,147.02) node [anchor=north west][inner sep=0.75pt]    {$V_{-}$};
\draw (121,97.02) node [anchor=north west][inner sep=0.75pt]    {$I_{-}$};
\draw (333,210.95) node    {$ $};
\draw (27.65,82.27) node   [align=left] {{\fontfamily{ptm}\selectfont \textit{U}-spin}};
\draw (225,82.27) node   [align=left] {{\fontfamily{ptm}\selectfont \textit{V}-spin}};
\draw (225,34.89) node    {$ $};
\draw (27.65,34.89) node    {$ $};
\draw (278.1,224.27) node   [align=left] {$\displaystyle I${\fontfamily{ptm}\selectfont -spin}};
\draw (231,210.95) node    {$\Sigma ^{+}$};
\draw (27.65,210.95) node    {$\Sigma ^{-}$};
\draw (78.49,122.89) node    {$n$};
\draw (180.16,299) node    {$\Xi ^{0}$};
\draw (78.49,299) node    {$\Xi ^{-}$};
\draw (180.16,122.89) node    {$p$};
\draw (129.32,210.95) node    {$\Sigma ^{0} /\Lambda $};
\draw    (246,210.95) -- (322,210.95) ;
\draw [shift={(324,210.95)}, rotate = 180] [color={rgb, 255:red, 0; green, 0; blue, 0 }  ][line width=0.75]    (10.93,-3.29) .. controls (6.95,-1.4) and (3.31,-0.3) .. (0,0) .. controls (3.31,0.3) and (6.95,1.4) .. (10.93,3.29)   ;
\draw    (186.28,110.89) -- (217.98,48.68) ;
\draw [shift={(218.89,46.89)}, rotate = 117] [color={rgb, 255:red, 0; green, 0; blue, 0 }  ][line width=0.75]    (10.93,-3.29) .. controls (6.95,-1.4) and (3.31,-0.3) .. (0,0) .. controls (3.31,0.3) and (6.95,1.4) .. (10.93,3.29)   ;
\draw    (71.56,110.89) -- (35.58,48.63) ;
\draw [shift={(34.58,46.89)}, rotate = 59.99] [color={rgb, 255:red, 0; green, 0; blue, 0 }  ][line width=0.75]    (10.93,-3.29) .. controls (6.95,-1.4) and (3.31,-0.3) .. (0,0) .. controls (3.31,0.3) and (6.95,1.4) .. (10.93,3.29)   ;
\draw    (216,210.95) -- (158.32,210.95) ;
\draw [shift={(155.32,210.95)}, rotate = 360] [fill={rgb, 255:red, 0; green, 0; blue, 0 }  ][line width=0.08]  [draw opacity=0] (8.93,-4.29) -- (0,0) -- (8.93,4.29) -- cycle    ;
\draw  [dash pattern={on 4.5pt off 4.5pt}]  (223.49,197.95) -- (187.09,134.89) ;
\draw    (189.17,283.4) -- (223.49,223.95) ;
\draw [shift={(187.67,286)}, rotate = 300] [fill={rgb, 255:red, 0; green, 0; blue, 0 }  ][line width=0.08]  [draw opacity=0] (8.93,-4.29) -- (0,0) -- (8.93,4.29) -- cycle    ;
\draw    (103.32,210.95) -- (45.15,210.95) ;
\draw [shift={(42.15,210.95)}, rotate = 360] [fill={rgb, 255:red, 0; green, 0; blue, 0 }  ][line width=0.08]  [draw opacity=0] (8.93,-4.29) -- (0,0) -- (8.93,4.29) -- cycle    ;
\draw  [dash pattern={on 4.5pt off 4.5pt}]  (35.16,223.95) -- (70.98,286) ;
\draw    (71.56,134.89) -- (36.66,195.35) ;
\draw [shift={(35.16,197.95)}, rotate = 300] [fill={rgb, 255:red, 0; green, 0; blue, 0 }  ][line width=0.08]  [draw opacity=0] (8.93,-4.29) -- (0,0) -- (8.93,4.29) -- cycle    ;
\draw    (171.16,122.89) -- (90.49,122.89) ;
\draw [shift={(87.49,122.89)}, rotate = 360] [fill={rgb, 255:red, 0; green, 0; blue, 0 }  ][line width=0.08]  [draw opacity=0] (8.93,-4.29) -- (0,0) -- (8.93,4.29) -- cycle    ;
\draw    (95.99,299) -- (166.16,299) ;
\draw [shift={(92.99,299)}, rotate = 0] [fill={rgb, 255:red, 0; green, 0; blue, 0 }  ][line width=0.08]  [draw opacity=0] (8.93,-4.29) -- (0,0) -- (8.93,4.29) -- cycle    ;
\draw    (87.49,283.4) -- (121.82,223.95) ;
\draw [shift={(85.99,286)}, rotate = 300] [fill={rgb, 255:red, 0; green, 0; blue, 0 }  ][line width=0.08]  [draw opacity=0] (8.93,-4.29) -- (0,0) -- (8.93,4.29) -- cycle    ;
\draw    (138.33,195.35) -- (173.23,134.89) ;
\draw [shift={(136.83,197.95)}, rotate = 300] [fill={rgb, 255:red, 0; green, 0; blue, 0 }  ][line width=0.08]  [draw opacity=0] (8.93,-4.29) -- (0,0) -- (8.93,4.29) -- cycle    ;

\end{tikzpicture}
}

\end{minipage}
    \caption{$I_{-},V_{-}$}
    \label{fig:swart_fail}
  \end{subfigure}
  \begin{subfigure}{0.3\textwidth}
    \centering
    \tikzset{every picture/.style={line width=0.75pt}} 

\begin{minipage}{\linewidth}
\centering

\scalebox{0.5}{

\begin{tikzpicture}[x=0.75pt,y=0.75pt,yscale=-1,xscale=1]


\draw (129.32,196.95) node    {$\Sigma ^{0} /\Lambda $};
\draw (180.16,108.89) node    {$p$};
\draw (78.49,285) node    {$\Xi ^{-}$};
\draw (180.16,285) node    {$\Xi ^{0}$};
\draw (78.49,108.89) node    {$n$};
\draw (27.65,196.95) node    {$\Sigma ^{-}$};
\draw (231,196.95) node    {$\Sigma ^{+}$};
\draw (278.1,210.27) node   [align=left] {$\displaystyle I${\fontfamily{ptm}\selectfont -spin}};
\draw (27.65,20.89) node    {$ $};
\draw (225,20.89) node    {$ $};
\draw (225,68.27) node   [align=left] {{\fontfamily{ptm}\selectfont \textit{V}-spin}};
\draw (27.65,68.27) node   [align=left] {{\fontfamily{ptm}\selectfont \textit{U}-spin}};
\draw (333,196.95) node    {$ $};
\draw (121,83.02) node [anchor=north west][inner sep=0.75pt]    {$I_{+}$};
\draw (30,134.02) node [anchor=north west][inner sep=0.75pt]    {$V_{-}$};
\draw (214,233.02) node [anchor=north west][inner sep=0.75pt]    {$V_{-}$};
\draw    (246,196.95) -- (322,196.95) ;
\draw [shift={(324,196.95)}, rotate = 180] [color={rgb, 255:red, 0; green, 0; blue, 0 }  ][line width=0.75]    (10.93,-3.29) .. controls (6.95,-1.4) and (3.31,-0.3) .. (0,0) .. controls (3.31,0.3) and (6.95,1.4) .. (10.93,3.29)   ;
\draw    (186.28,96.89) -- (217.98,34.68) ;
\draw [shift={(218.89,32.89)}, rotate = 117] [color={rgb, 255:red, 0; green, 0; blue, 0 }  ][line width=0.75]    (10.93,-3.29) .. controls (6.95,-1.4) and (3.31,-0.3) .. (0,0) .. controls (3.31,0.3) and (6.95,1.4) .. (10.93,3.29)   ;
\draw    (71.56,96.89) -- (35.58,34.63) ;
\draw [shift={(34.58,32.89)}, rotate = 59.99] [color={rgb, 255:red, 0; green, 0; blue, 0 }  ][line width=0.75]    (10.93,-3.29) .. controls (6.95,-1.4) and (3.31,-0.3) .. (0,0) .. controls (3.31,0.3) and (6.95,1.4) .. (10.93,3.29)   ;
\draw    (189.17,269.4) -- (223.49,209.95) ;
\draw [shift={(187.67,272)}, rotate = 300] [fill={rgb, 255:red, 0; green, 0; blue, 0 }  ][line width=0.08]  [draw opacity=0] (8.93,-4.29) -- (0,0) -- (8.93,4.29) -- cycle    ;
\draw  [dash pattern={on 4.5pt off 4.5pt}]  (223.49,183.95) -- (187.09,120.89) ;
\draw    (213,196.95) -- (155.32,196.95) ;
\draw [shift={(216,196.95)}, rotate = 180] [fill={rgb, 255:red, 0; green, 0; blue, 0 }  ][line width=0.08]  [draw opacity=0] (8.93,-4.29) -- (0,0) -- (8.93,4.29) -- cycle    ;
\draw    (71.56,120.89) -- (36.66,181.35) ;
\draw [shift={(35.16,183.95)}, rotate = 300] [fill={rgb, 255:red, 0; green, 0; blue, 0 }  ][line width=0.08]  [draw opacity=0] (8.93,-4.29) -- (0,0) -- (8.93,4.29) -- cycle    ;
\draw  [dash pattern={on 4.5pt off 4.5pt}]  (35.16,209.95) -- (70.98,272) ;
\draw    (100.32,196.95) -- (42.15,196.95) ;
\draw [shift={(103.32,196.95)}, rotate = 180] [fill={rgb, 255:red, 0; green, 0; blue, 0 }  ][line width=0.08]  [draw opacity=0] (8.93,-4.29) -- (0,0) -- (8.93,4.29) -- cycle    ;
\draw    (168.16,108.89) -- (87.49,108.89) ;
\draw [shift={(171.16,108.89)}, rotate = 180] [fill={rgb, 255:red, 0; green, 0; blue, 0 }  ][line width=0.08]  [draw opacity=0] (8.93,-4.29) -- (0,0) -- (8.93,4.29) -- cycle    ;
\draw    (92.99,285) -- (163.16,285) ;
\draw [shift={(166.16,285)}, rotate = 180] [fill={rgb, 255:red, 0; green, 0; blue, 0 }  ][line width=0.08]  [draw opacity=0] (8.93,-4.29) -- (0,0) -- (8.93,4.29) -- cycle    ;
\draw    (87.49,269.4) -- (121.82,209.95) ;
\draw [shift={(85.99,272)}, rotate = 300] [fill={rgb, 255:red, 0; green, 0; blue, 0 }  ][line width=0.08]  [draw opacity=0] (8.93,-4.29) -- (0,0) -- (8.93,4.29) -- cycle    ;
\draw    (138.33,181.35) -- (173.23,120.89) ;
\draw [shift={(136.83,183.95)}, rotate = 300] [fill={rgb, 255:red, 0; green, 0; blue, 0 }  ][line width=0.08]  [draw opacity=0] (8.93,-4.29) -- (0,0) -- (8.93,4.29) -- cycle    ;

\end{tikzpicture}

}

\end{minipage}
    \caption{$I_{+},V_{-}$}
    \label{fig:swart_success}
  \end{subfigure}
  \begin{subfigure}{0.3\textwidth}
    \centering
    \tikzset{every picture/.style={line width=0.75pt}} 

\begin{minipage}{\linewidth}
\centering

\scalebox{0.5}{

\begin{tikzpicture}[x=0.75pt,y=0.75pt,yscale=-1,xscale=1]

\draw  (275,72.55) -- (317.55,72.55)(275,30) -- (275,72.55) -- cycle (310.55,67.55) -- (317.55,72.55) -- (310.55,77.55) (270,37) -- (275,30) -- (280,37)  ;

\draw (326,62.4) node [anchor=north west][inner sep=0.75pt]    {$I_{3}$};
\draw (266,12.4) node [anchor=north west][inner sep=0.75pt]    {$Y$};
\draw (135.32,201.32) node    {$\Sigma ^{0} /\Lambda $};
\draw (186.16,113.27) node    {$p$};
\draw (84.49,289.38) node    {$\Xi ^{-}$};
\draw (186.16,289.38) node    {$\Xi ^{0}$};
\draw (84.49,113.27) node    {$n$};
\draw (33.65,201.32) node    {$\Sigma ^{-}$};
\draw (237,201.32) node    {$\Sigma ^{+}$};
\draw (33.65,25.27) node    {$ $};
\draw (231,25.27) node    {$ $};
\draw (336.5,201.5) node    {$ $};
\draw (128,92.4) node [anchor=north west][inner sep=0.75pt]    {$I_{-}$};
\draw (210,132.4) node [anchor=north west][inner sep=0.75pt]    {$U_{-}$};
\draw (285,214) node   [align=left] {$\displaystyle I${\fontfamily{ptm}\selectfont -spin}};
\draw (32.5,73.5) node   [align=left] {{\fontfamily{ptm}\selectfont \textit{U}-spin}};
\draw (233.5,73.5) node   [align=left] {{\fontfamily{ptm}\selectfont \textit{V}-spin}};
\draw    (252,201.35) -- (325.5,201.48) ;
\draw [shift={(327.5,201.48)}, rotate = 180.1] [color={rgb, 255:red, 0; green, 0; blue, 0 }  ][line width=0.75]    (10.93,-3.29) .. controls (6.95,-1.4) and (3.31,-0.3) .. (0,0) .. controls (3.31,0.3) and (6.95,1.4) .. (10.93,3.29)   ;
\draw    (192.28,101.27) -- (223.98,39.05) ;
\draw [shift={(224.89,37.27)}, rotate = 117] [color={rgb, 255:red, 0; green, 0; blue, 0 }  ][line width=0.75]    (10.93,-3.29) .. controls (6.95,-1.4) and (3.31,-0.3) .. (0,0) .. controls (3.31,0.3) and (6.95,1.4) .. (10.93,3.29)   ;
\draw    (77.56,101.27) -- (41.58,39) ;
\draw [shift={(40.58,37.27)}, rotate = 59.99] [color={rgb, 255:red, 0; green, 0; blue, 0 }  ][line width=0.75]    (10.93,-3.29) .. controls (6.95,-1.4) and (3.31,-0.3) .. (0,0) .. controls (3.31,0.3) and (6.95,1.4) .. (10.93,3.29)   ;
\draw  [dash pattern={on 4.5pt off 4.5pt}]  (193.67,276.38) -- (229.49,214.32) ;
\draw    (227.99,185.73) -- (193.09,125.27) ;
\draw [shift={(229.49,188.32)}, rotate = 240] [fill={rgb, 255:red, 0; green, 0; blue, 0 }  ][line width=0.08]  [draw opacity=0] (8.93,-4.29) -- (0,0) -- (8.93,4.29) -- cycle    ;
\draw    (222,201.32) -- (164.32,201.32) ;
\draw [shift={(161.32,201.32)}, rotate = 360] [fill={rgb, 255:red, 0; green, 0; blue, 0 }  ][line width=0.08]  [draw opacity=0] (8.93,-4.29) -- (0,0) -- (8.93,4.29) -- cycle    ;
\draw  [dash pattern={on 4.5pt off 4.5pt}]  (77.56,125.27) -- (41.16,188.32) ;
\draw    (41.16,214.32) -- (75.48,273.78) ;
\draw [shift={(76.98,276.38)}, rotate = 240] [fill={rgb, 255:red, 0; green, 0; blue, 0 }  ][line width=0.08]  [draw opacity=0] (8.93,-4.29) -- (0,0) -- (8.93,4.29) -- cycle    ;
\draw    (109.32,201.32) -- (51.15,201.32) ;
\draw [shift={(48.15,201.32)}, rotate = 360] [fill={rgb, 255:red, 0; green, 0; blue, 0 }  ][line width=0.08]  [draw opacity=0] (8.93,-4.29) -- (0,0) -- (8.93,4.29) -- cycle    ;
\draw    (126.32,185.73) -- (91.42,125.27) ;
\draw [shift={(127.82,188.32)}, rotate = 240] [fill={rgb, 255:red, 0; green, 0; blue, 0 }  ][line width=0.08]  [draw opacity=0] (8.93,-4.29) -- (0,0) -- (8.93,4.29) -- cycle    ;
\draw    (177.16,113.27) -- (96.49,113.27) ;
\draw [shift={(93.49,113.27)}, rotate = 360] [fill={rgb, 255:red, 0; green, 0; blue, 0 }  ][line width=0.08]  [draw opacity=0] (8.93,-4.29) -- (0,0) -- (8.93,4.29) -- cycle    ;
\draw    (177.16,273.78) -- (142.83,214.32) ;
\draw [shift={(178.66,276.38)}, rotate = 240] [fill={rgb, 255:red, 0; green, 0; blue, 0 }  ][line width=0.08]  [draw opacity=0] (8.93,-4.29) -- (0,0) -- (8.93,4.29) -- cycle    ;
\draw    (101.99,289.38) -- (172.16,289.38) ;
\draw [shift={(98.99,289.38)}, rotate = 0] [fill={rgb, 255:red, 0; green, 0; blue, 0 }  ][line width=0.08]  [draw opacity=0] (8.93,-4.29) -- (0,0) -- (8.93,4.29) -- cycle    ;

        

\end{tikzpicture}
}

\end{minipage}
    \caption{$I_{-},U_{-}$}
    \label{fig:chen_IU}
  \end{subfigure}  
  \caption{Tracks of ladder operators on octet baryons.}
  \label{fig:ladderCompare}
\end{figure}
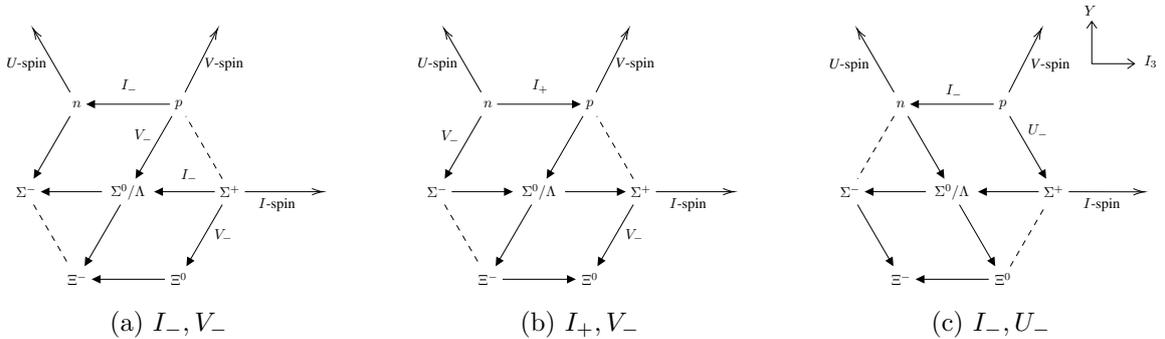

It is expected that, getting the CGCs for $SU_3$ are more involved than $SU_2$.
The main reason originates from the fact that the rank of $SU_3$ is two, which requires two descending operators (instead of one $J_-$ in $SU_2$).
In $SU_3$, we have three descending operators to select from, $I_-, U_-, V_-$ (see Figure~\ref{fig:ladderCompare}).

Based on the experience of $SU_2$, we intend to keep the operator $I_-$ to be one of two descending operators,
otherwise it would be a restart instead of an extension of $SU_2$.
This choice also has a physical reason that we can easily track different isospin multiplets.
One may want to make an assumption that the matrix elements of $I_\pm, U_\pm, V_\pm$ can be tuned to be positive, however, the three operators cannot be simultaneously positive due to the structure of the $SU_3$ Lie algebra\footnote{
See, e.g., $U_-$ matrix in $(p,q)=(1,1)$ representation in Appendix.~\eqref{sec:wf_swart}.
}.

One may speculate that selecting $\{I_\pm, V_\pm\}$ is the same as $\{I_\pm, U_\pm\}$, however, 
we will show there is a mathematical reason that the latter selection is superior.

From subplot~\eqref{fig:swart_fail} of Figure~\ref{fig:ladderCompare}, we learn that to enumerate all the states in the root space (or the weight space of the adjoint representation) with only descending operators $I_{-}, V_{-}$, one has to start from two "highest" state $p, \Sigma^+$.
The consequence is that one cannot naturally define the highest weight.
To enumerate all of the octet, 
we need both descending operator and the ascending operator, i.e., 
by $I_{+},V_{-}$ from $n$(see Figure~\ref{fig:swart_success}), or $I_{-}, V_{+}$ starting from $\Xi^0$.
However, both "highest/lowest" starting weights are unconventional and counter-intuitive.
Despite that nothing stops one to assign an additional convention to the order of the octet states, 
this extra convention is essentially unnecessary.

In contrast, the convention of choosing $\{I_{\pm}, U_{\pm}\}$ to be positive is free from this dilemma.
All the weights within any representations can be enumerated by pure descending operators $\{I_{-}, U_{-}\}$.
This fact can be easily seen by noting that the angle between  $I_{-}, U_{-}$ is $120^\circ$, and this obtuse angle makes the operator pair be capable of enumerating all the weight vectors in any representation, especially in the case like the octet, where the envelope polygon has obtuse angles\footnote{This is also the reason why $I_{+}, V_{-}$ or $I_{-}, V_{+}$ can also do the job, 
but, as we have pointed out, if one operator is ascending operator, it will bring ambiguity to the choice of a highest weight.}.

To conclude, as long as one keeps the selection of $I_-$ operator, the positive $\{I_-, U_-\}$ operator set is the only way to naturally extend the $J_-$ operator in $SU_{2}$.

The second task is to fix the sign of $\ket{I_1, I_{z1},Y_1; I_2, I_{z2},Y_2}$ in the highest weight.
Haacke et.al.~\cite{Haacke:1975rt}, de Swart~\cite{deSwart:1963pdg} and Rabl et.al.~\cite{Rabl:1975zy} 
all take essentially the same convention as $SU_2$,
i.e., in the $SU_3$ CGCs, the largest isospin of the first particle $I_1$ of the highest weight is assumed to be positive.

However, note that the essence of the second step is to define an order for the uncoupled representation, since $\{I_-, U_-\}$ already defines a natural order for all the multiplets, the extra assignment of the order is essentially unnecessary.
Thus, we extend the $SU_2$ Condon-Shortley convention to the requirement that,
in the $SU_N$ CGCs, the coefficient of the highest weight (instead of the largest isospin $I$) of the first particle is positive \footnote{In $SU_2$, the highest weights happens to be the largest isospin $I$ (or angular momentum $J$).}.
We call this convention the generalized Condon-Shortley convention.

It is reasonable to speculate that different conventions will lead to different CGCs and isoscalar factors (ISFs)\ref{sec:isf}.
This turns out to be the case, and we will give the detailed discussion in Section~\eqref{sec:isf}.

Our definition of the order for the $SU_3$ multiplets will be well-defined in the non-degenerate case.
However, in some degenerate cases, such as $8 \otimes 8 = 27\oplus 10\oplus 10^*\oplus 8\oplus 8\oplus 1$, where the octet occurs twice, any rotation between the two octets is a valid CGCs.
This degeneracy can only be broken by additional symmetry,
and it is conventional to demand that the CGCs of the $8\times 8$ is split into symmetric and anti-symmetric part.
Here we use the same convention as that of Ref.~\cite{Haacke:1975rt, Rabl:1975zy}, namely, the symmetric one is superior to the anti-symmetric one.

At this stage, all the mathematics of the CGC is settled.
Once the matrix elements of the operators is given 
(see, e.g., equation~(48) in Ref.~\cite{Baird:1963wv} or equation~(3.3) in Ref.~\cite{Haacke:1975rt}).
We can repeat and extend the process in $SU_2$, 
which includes recursively applying the descending operators and doing the orthogonalization with predefined phase conventions.

In principle, it is not difficult to turn these rules into computer programs.
However, it is worthy to mention that, this method is still cumbersome in practice and not very efficient to generalize to larger groups.
The eigen function method (EFM) invented by Jin-Quan Chen et.al.~\cite{Chen:1985zzd, Chen:2002gd} solves this problem once for all.
After a delicate construction of the complete set of commuting operators and and conventions of the eigenvector phases, EFM can yield the so called Gel'fand basis, which furnishes the irreducible basis of 
$SU_{n} \supset SU_{n-1}\otimes U_{1} \supset \ldots \supset SU_{2}\otimes U_{1} \supset U_{1}$.
Interested readers are referred to the monograph~\cite{Chen:2002gd}.

\subsection{Beyond Flavor $SU_3$}

Thing becomes more interesting when we push the flavor $SU_3$ symmetry to $SU_4$.
Although the flavor $SU_4$ symmetry is strongly broken by the heavy charm quark, it is worthwhile to study some mathematical properties.
Perhaps one unexpected result is that there is no baryon matrix in flavor $SU_4$.
This is a direct consequence of interplay between flavor and color symmetry.

In the previous sections, we only focus on the flavor symmetry.
It is time to talk about the color symmetry.
Unlike the flavor symmetry, which is only fulfilled by the hadrons approximately, the color symmetry is an exact one.

The fundamental theory of the strong interactions is the quantum chromodynamics (QCD), which is a $SU_3$ gauge theory on the color degree of freedom.
So far, all the observed hadrons observed are color-singlet or colorless.
Although not theoretically proved, it is widely believed that colors are constrained within hadrons and all hadrons should be colorless.
This is an important and stringent constrain.

For baryons, the only way to get the color singlet is $3n$ quarks with possible quark-antiquark pairs, where $n$ is the baryon number of system.
Formally, we can continue the trick of trading one antiquark with two quarks, so the color-singlet requirement always means $3n$ quarks.

For conventional baryons, with 3 quarks at disposal, we have the following tensor decomposition in the \textit{flavor} degree of freedom,
\begin{align}
\yng(1) \otimes \yng(1) \otimes \yng(1) &= \yng(3) \oplus \yng(2,1) \oplus \yng(2,1) \oplus \yng(1,1,1), \\
3 \otimes 3 \otimes 3 &= 10 \oplus 8 \oplus 8 \oplus 1, \mathrm{for \ } SU_3 ,\\
4 \otimes 4 \otimes 4 &= 20 \oplus 20 \oplus 20 \oplus \bar{4}, \mathrm{for \ } SU_4 . \label{eqn:su4baryon}
\end{align}

For flavor $SU_4$, we have,
\begin{align}
\yng(1) \otimes \yng(1,1,1) &= \yng(2,1,1) \oplus \yng(1,1,1,1), \\
4 \otimes \bar{4} &= 15 \oplus 1.
\end{align}

The adjoint representation of $SU_3$ and $SU_4$ are the irreps of $\yng(2,1)$ and $\yng(2,1,1)$, respectively.
The adjoint rep shows up naturally as a result of tensor product of fundamental and complex conjugate representation.
For $SU_3$, the motivation of constructing the matrix form of octet baryons and mesons is to reveal explicitly the decomposition process.

Namely, $M \to U M U^\dagger$, where $U$ is the transformation matrix in fundamental representation and $M$ is the octet baryon or meson matrix.
For other irreps, such as $10$ decuplet in $SU_3$ flavor symmetry, one would have to construct explicitly the $10 \times 10$ matrix for each generator.
In this case, $D \to R_{10\times 10}(U) D$, 
and the decuplet baryon is a column vector, 
they cannot be organized into a $3\times 3$ matrix form as the adjoint representation.
In practice, however, this 10 dimensional vector is rarely used.
Instead, people group them into different isospin multiplets and treat separately.
Essentially, the matrix and vector forms of the hadrons are nothing but convenient realizations of the underline CGCs.

From the Eqn.~\eqref{eqn:su4baryon}, we can see that adjoint $\yng(2,1,1)$ does not show up in the decomposition.
There is no such thing like $SU_4$ baryon matrix, only meson matrix is possible.
It is a lucky coincidence that the flavor $SU_3$ symmetry happens to be the same as color $SU_{3}$ symmetry.

\section{Group Theory from the Quark Model}\label{sec:quark}

\subsection{Antiquarks and the Complex Conjugation Representation}

From the perspective of the quark model, hadrons are made up of quarks and anti-quarks.
The quark is assumed to furnish the fundamental representation of $SU_3$\footnote{
Here, we focus on the quarks with three flavors instead of six.
This setting is extensively studied in the literature.}.
A straightforward definition of an anti-quark is that it resides in the complex conjugate representation of the fundamental representation, denoted as $3^*$.
This definition has the advantage that the singlet $\mathbbm{1}$ has a easy form,
\begin{align}
    \mathbbm{1} &\propto u u^* + d d^* + s s^* = q^i q^{i*},\\
    \xrightarrow{SU_3} &= U^i_{\ j} q^j (U^i_{\ k})^* q^{k*}\\
    &= U^i_{\ j} (U^*)^i_{\ k} q^j q^{k*}\\
    &= U^i_{\ j} (U^\dagger)^k_{\ i} q^j q^{k*} = (U^\dagger U)^k_j q^j q^{k*} \\
    &= \delta_j^k q^j q^{k*} = q^i q^{i*},
\end{align}
where $U$ is the fundamental representation matrix.
To further simplify the notation, it is conventional to group the antiquarks into a \textit{row} vector, whose transformation property can be compactly written as
\begin{align}
    \bar{q}:= (q^*)^T \Rightarrow \bar{q}' &= \bar{q} U^\dagger \equiv \bar{q} U^{-1}.
\end{align}
Specifically, for flavor $SU_3$, we have 
\begin{align}
    \triplet{u'^*}{d'^*}{s'^*} = U^* \triplet{u^*}{d^*}{s^*} &\Leftrightarrow (\bar{u}',\bar{d}',\bar{s}') = (\bar{u},\bar{d},\bar{s})U^\dagger.\label{eqn:antiquark}
\end{align}
Then the adjoint representation $M$ has a natural transformation property $M \to UMU^\dagger$.
This property is extensively used to simplify the construction process of the Lagrangian in the chiral perturbation theory (ChPT). 

This Hermitian conjugate also leads to a readily decomposition for $3\otimes\bar{3}$ as follows,
\begin{align}
\yng(1) \otimes \yng(1,1) &= \yng(2,1) \oplus \yng(1,1,1), \\
3 \otimes \bar{3} &= 8 \oplus 1, \\
\begin{tabular}{cccc}
  & $\bar{u}$ & $\bar{d}$ & $\bar{s}$\\
$u$ & \multirow{3}{*}{$\left( 
    \begin{array}{c}
         \\
         \\
         \\ 
    \end{array}
\right.$} & & \multirow{3}{*}{$\left. 
    \begin{array}{c}
        \\
        \\
        \\ 
    \end{array}
\right)$} \\
$d$ & & & \\
$s$ & & & \\
\end{tabular} &= 
\left(
    \arraycolsep1pt
\begin{array}{ccc}
    \frac{1}{3}(2u\bar{u}-d\bar{d}-s\bar{s}) & u\bar{d} & u\bar{s}\\
    d\bar{u} & \frac{1}{3}(2d\bar{d}-u\bar{u}-s\bar{s}) & d\bar{s} \\
    s\bar{u} & s\bar{d} & \frac{1}{3}(2s\bar{s}-u\bar{u}-d\bar{d})\\
\end{array}
\right) \nonumber\\
&+ \frac{1}{3}(u\bar{u}+d\bar{d}+s\bar{s}) \mathbbm{1}_{3\times 3}. \label{eqn:su3decompose}
\end{align}
Identifying the decomposition on the right hand side of Eqn.~\eqref{eqn:su3decompose} with hadrons is equivalent to specifying the hadron flavor wave functions.
This is the process of adopting a \textbf{hadron flavor convention}.

We need to point it out that, although, in principle, one can adopt a different phase convention for the anti-quarks, with the consequence that his/her adjoint representation $M$ would transform differently from usual $UMU^\dagger$.
For example, one can define his/her $\bar{s}$ to be the negative of our $\bar{s}$, then his/her singlet would be proportional to $u \bar{u} + d \bar{d} - s \bar{s}$, 
which is quite bizarre and counter-intuitive.
In practice, it would cause confusions and, 
in worst cases, misinterpretations of the intermediate steps by other phenomenological model, like ChPT.
Since the difference is just trivially notational without any profound reason, we see no need to invent a new convention for the anti-quarks.

It is worthy to pointed out that, in the famous paper \cite{deSwart:1963pdg} by de Swart \footnote{His work shows up just before the dawn of the quark model.},
redefinition of possible phases are involved to maintain the positivity of $I_\pm, V_\pm$ matrix elements in any representation $(p,q)$.
With the language of modern quark model, $p,q$ represents the number of quarks and anti-quarks, respectively, 
and his requirement can be boiled down to the phase redefinition of the anti-quarks.

Equation (8.2) in Ref.~\cite{deSwart:1963pdg} can be extend to manage fractional-charged quarks
\begin{align}
    q'^*\equiv\phi(\{N^*\}, I_z, Y) := \eta (-)^{I_z -\frac{3}{2}Y} \phi^*(\{N\}, -I_z, -Y) \equiv\eta (-)^{I_z -\frac{3}{2}Y}q^*.
    \label{eqn:swart_extend}
\end{align}
For $q=u,d,s$ quarks, this would result in 
\begin{align}
    u'^* = \eta u^*, d'^* = - \eta d^*, s'^* = - \eta s^*.
\end{align}

In contrast to the physical particles where $(I_z, Y)= (0,0)$ always shows up in any irrep $(p,q)$, one can naturally fix the $\eta=1$ by requiring that $\phi(\{N^*\}, 0, 0) := \phi^*(\{N\}, 0,0)$ (c.f. equation (8.3) in Ref.~\cite{deSwart:1963pdg}).
There is no additional natural phases to pinpoint the phase $\eta$ in Eqn.~\eqref{eqn:swart_extend} at the quark level.
If, for whatever reason, $\eta=1$, then the singlet would be 
\begin{align}
    uu'^* - dd'^* - ss'^* = uu^* + dd^* + ss^*,
\end{align}
where the additional negative sign on the left hand side is due to the CGCs of $3\otimes \bar{3} \to 1$ under this convention.
We do not adopt this additional redefinition of the anti-quarks due to the reason we explained above.

\subsection{Antiquarks and the Isospin Convention}\label{sec:iso_con}

There is another way to represent the antiquarks from the anti-symmetrized combination of quarks.
This way will also leads to the definition of the isospin convention.

To start, recall that for a system consisting of $m$ particles with each particle furnishes a representation of a group, the total wave function is a tensor product of each degree of freedom,
\begin{align}
    (\psi')^{i_1, i_2,\ldots,i_m} = D(R)^{i_1}_{\ j_1} D(R)^{i_1}_{\ j_1} \ldots D(R)^{i_1}_{\ j_1} \phi^{j_1} \phi^{j_2}\ldots \phi^{j_m},
\end{align}
where $D(R)$ is the representation matrix of a group.
Note the following mathematical fact,
\begin{align}
\epsilon_{\Range{i}{N}}A^{i_1}_{\ j_1}A^{i_2}_{\ j_2}\ldots A^{i_N}_{\ j_N}&=\epsilon_{\Range{j}{N}}\mathrm{det}(A)
\end{align}
where $A$ is an arbitrary square matrix and $\epsilon_{\Range{i}{N}}$ is the Levi-Civita symbol.
Replace $A$ with unitary matrix $U$, we have 
\begin{align}
\epsilon_{\Range{i}{N}}U^{i_1}_{\ j_1}U^{i_2}_{\ j_2}\ldots U^{i_N}_{\ j_N}&=\epsilon_{\Range{j}{N}}. \label{eqn:epsilon}
\end{align}
From the above equation, we can define a $SU_n$ singlet by
\begin{align}
\mathbbm{1} &=\frac{1}{\sqrt{N!}}\epsilon_{\Range{i}{N}}\psi^{\Range{i}{N}}, \label{eqn:siglet}\\
\psi^{\Range{i}{N}} &:= \psi^{i_1}\psi^{i_2}\ldots\psi^{i_N},
\end{align}
where $1/\sqrt{N!}$ is the normalization constant.
This statement can be checked by
\begin{align}
\mathbbm{1}' &=\frac{1}{\sqrt{N!}}\epsilon_{\Range{i}{N}}U^{i_1}_{\ j_1}U^{i_2}_{\ j_2}\ldots U^{i_N}_{\ j_N}\psi^{\Range{j}{N}},\\
    &= \frac{1}{\sqrt{N!}}\epsilon_{\Range{j}{N}}\psi^{\Range{j}{N}},\\
    &= \mathbbm{1}.
\end{align}

On the other hand, contracting with $(U^\dagger)^{j_1}_{\ k_1}$ on both sides of Eqn.~\eqref{eqn:epsilon}, 
we will arrive at,
\begin{align}
\epsilon_{\Range{i}{N}}(U^{i_1}_{\ j_1} (U^\dagger)^{j_1}_{\ k_1})U^{i_2}_{\ j_2}\ldots U^{i_N}_{\ j_N}&=\epsilon_{\Range{j}{N}} (U^\dagger)^{j_1}_{\ k_1}\,, \\
\epsilon_{\Range{i}{N}} U^{i_2}_{\ j_2}\ldots U^{i_N}_{\ j_N}&=\epsilon_{\Range{j}{N} } (U^\dagger)^{j_1}_{\ i_1}\,.
\end{align}

Inspired by this equation and singlet state as shown in Eqn.~\eqref{eqn:siglet}, we can define a new state $\psi_{i_1}$ as,
\begin{align}
\psi_{i_1} := \frac{1}{\sqrt{(N-1)!}}\epsilon_{\Range{i}{N}} \psi^{i_2, i_3, \ldots, i_N} \label{eqn:qbar_def}
\end{align}
i.e., instead of contracting all the indices of the Levi-Civita tensor, we choose to keep the first index $i_1$.
This new state transformers as,
\begin{align}
\psi'_{i_1} &= \frac{1}{\sqrt{(N-1)!}} \epsilon_{\Range{i}{N}} U^{i_2}_{\ j_2}U^{i_3}_{\ j_3}\ldots U^{i_N}_{\ j_N} \psi^{j_2, j_3, \ldots, j_N}\\
&= \frac{1}{\sqrt{(N-1)!}} \epsilon_{\Range{j}{N} } \psi^{j_2, j_3, \ldots, j_N} (U^\dagger)^{j_1}_{\ i_1} \\
&= \psi_{j_1} (U^\dagger)^{j_1}_{\ i_1} = (U^*)_{i_1}^{\ j_1} \psi_{j_1},
\end{align}
where in the last step $(A^T)_i^{\ j} = A^j_{\ i}$ is used\footnote{The order of the matrix indices represent the row-column relation, and in cases where only quarks are involved, one can safely write only with lower indices.}.
So $\psi_{i}$ transforms as the complex conjugate representation, and we call it anti-quark in the context of group theory.
The last step also tells us that, in $SU_n$, the complex conjugate representation is equivalent to applying $U^\dagger$ from the right side, i.e., $\bar{q}' = \bar{q} U^\dagger = \bar{q} U^{-1}$.

We need to stress that, we have kept the \textbf{first} index free in the definition \eqref{eqn:qbar_def} of antiquarks.
However, in principle, one can pick any free index in $\Range{i}{N}$, and by choosing a specific one, one pick a specific phase convention for antiquarks.
Notably, in the special case of the isospin symmetry which belongs to $SU_2$ group,
one can let the first index to be free as we do,
\begin{align}
\bar{u} &:= \epsilon_{1,2} d = d\,,\\
\bar{d} &:= \epsilon_{2,1} u = -u\,,
\end{align}
or choose to keep the last index free as some authors do (Ref.\cite{Zee:group}),
\begin{align}
\bar{u} &:= \epsilon_{2,1} d = -d\,,\\
\bar{d} &:= \epsilon_{1,2} u = u\,.
\end{align}
The two choices will result in different conventions.
This convention is so important in hadron physics and we call it \textbf{isospin convention},
%
because in the strong interaction, the isospin symmetry is decently conserved, and the hadrons are conventionally organized into different isospin multiplets.

As can be seen from Eqn.~\eqref{eqn:qbar_def}, group
representing one antiquark by one quark is a property specific to $SU_2$ group, i.e., complex representation of $SU_2$ group can be achieved by a linear transformation of the fundamental representation $\doublet{u}{d}$.
This tells us that $SU_2$ group has no complex representation (only pseudo-real/quaternionic representation).

Exchanging an anti-quark with the anti-symmetrized $N-1$ quarks is reminiscent of the Dirac sea.
This quark-antiquark duality is proved to be extremely useful in deriving the $SU_N$ CGCs~\cite{Chen:2002gd}.

\subsection{Convention Comparison}

Now we are ready to study the convention in de Swart's paper from perspective of the quark model.
As explained before, the states within a multiplet are linked by the descending operators, whose matrix elements are conventionally set to be positive.
To start, we should fix the phase of highest wave function, and with the hindsight of the quark model,
we set the first state in the octet to be 
\begin{align}
    \ket{8^{[1]}} := \ket{us^*}.
\end{align}
The second highest state can be got by $I_-$, i.e.,
\begin{align}
    I_- \ket{8^{[1]}} &= I_- \ket{us^*},\\
    \ket{8^{[2]}} &= \ket{(I_- u)s^*} + \ket{u (-(I_+)s)^*},\\
    &= \ket{ds^*}+ 0 = \ket{ds^*},
\end{align}
where $(I_\pm)^* = -(I_\mp)$ is used\footnote{
This complex conjugate here is what de Swart called the $\phi'$ representation in Ref.~\cite{deSwart:1963pdg}.
}.
To get the $\ket{8^{[3]}}$, 
we have to use the operator $U_-=[V_-,I_+]$.
Here we want to emphasize that the appearance of $I_+$ breaks the "descending" convention, 
and it also brings the ambiguity to the definition of the "highest" weight.

Applying $U_-$ to $\ket{8^{[1]}}$, we get, 
\begin{align}
    U_-\ket{8^{[1]}} &= U_- \ket{us^*},\\
    - \ket{8^{[3]}} &= \ket{(U_-u)s^*} + \ket{u (-(U_+)s)^*},\\
     &= 0+ \ket{u (-d)^*},\\
     \ket{8^{[3]}} &= \ket{ud^*}.
\end{align}
Please note the negative sign before $\ket{8^{[3]}}$ in the second line.
It is due to non-positiveness of $U_-$ in this convention.
The applications of $I_-, V_-$ on the rest states are straightforward, and we put the detailed steps in Appendix~\ref{sec:wf_swart}.

For comparison, we also list the octet states with the convention of choosing descending operator set $\{I_-, U_-\}$, which is used by Baird-Biedenharn~\cite{Biedenharn:1963zk, Baird:1963wv, Baird:1964zm, Baird:1964zn}, Haacke et.al.~\cite{Haacke:1975rt}, Rabl et.al.~\cite{Rabl:1975zy} and Chen et.al.~\cite{Chen:1985zzd, Chen:2002gd}.
We get the octet states with the language of the quark model as shown in Table~\ref{tab:IUV_m_compare}. 

\begin{table}[!htbp]
\centering
\setlength{\tabcolsep}{3pt}
\renewcommand{\arraystretch}{1.5}
\begin{tabular}{C|CCCCCCCC}
     & T_1 & T_2 & T_3 & T_4 & T_5 & T_6 & T_7 & T_8 \\
    \hline
    \{I_\pm, V_\pm\} & us^* & ds^* & ud^* & -\frac{1}{\sqrt{2}} (uu^* - dd^*) & -du^* & -\frac{1}{\sqrt{6}}(uu^* +dd^* -2ss^*) & sd^* & -su^*\\
    \{I_\pm, U_\pm\} & us^* & ds^* & -ud^* & \frac{1}{\sqrt{2}} (uu^* - dd^*) & du^* & -\frac{1}{\sqrt{6}}(uu^* +dd^* -2ss^*) & -sd^* & su^*\\
    \hline
    \text{de Swart\cite{deSwart:1963pdg}} & K^+ & K^0 & -\pi^+ & \pi^0 & \pi^- & \eta_8 & \bar{K}^0 & -K^-\\
    \text{Chen et.al.\cite{Chen:1979qz}} & K^+ & K^0 & \pi^+ & \pi^0 & \pi^- & \eta_8 & \bar{K}^0 & K^-\\
    \text{Rabl et.al.\cite{Rabl:1975zy}} & K^+ & K^0 & -\pi^+ & \pi^0 & \pi^- & -\eta_8 & -\bar{K}^0 & K^-\\
    \hline
\end{tabular}
    \caption{Flavor wave functions of the octet states, where $u^*, d^*,s^*$ can be identified with $\bar{u}, \bar{d},\bar{s}$ respectively, along with an additional transpose operation. See Eqn.\eqref{eqn:antiquark} in the main text. The last three rows are the hadron flavor convention, the convention from de Swart should be combined with the results from $\{I_\pm, V_\pm\}$; Chen and Rabl should be combined with that of $\{I_\pm, U_\pm\}$.}
    \label{tab:IUV_m_compare}
\end{table}

In Table~\ref{tab:IUV_m_compare}, 
$T_i, i=1,\ldots,8$ serve as the basis of the octet representation under different conventions.
Although octet mesons also serve as the basis of the octet, we can freely pick any phase conventions of their flavor wave functions, which we call the hadron flavor convention.
This kind of convention is also purely notational and thus it is independent of any mathematical deduction.
For instance, $\pi^+$ can be set freely to be $\pm u\bar{d}$ whether we choose operator set $\{I_\pm, V_\pm\}$ or $\{I_\pm, U_\pm\}$.

This flavor convention can only be fixed by conventions from physics.
One important consideration is the charge conjugation, e.g., if $K^+$ flavor wave function is chosen to be $u\bar{s}$, it is natural to assume that the wave function of its charge conjugate partner $K^-$ is $s\bar{u}$\footnote{ Here we shift the notation $q^*$ into $\bar{q}$, in order to be consistent with the notation in the modern quark model.}.
Since we conclude that the 8th basis $T_8$ is $-\bar{u}s$,
then $-K^-$ instead of $K^-$ should be identified with $T_8$.
Likewise, there is a relative negative sign between the wave functions of $T_3= u\bar{d}$ and $T_5= -d\bar{u}$,
one could assign $\pi^+= T_3=u\bar{d}, \pi^-= -T_5= d\bar{u}$ to eliminate the negative sign in the wave functions.
However, de Swart picked a different flavor convention, 
i.e., $\pi^+= -T_3=-u\bar{d}, \pi^-=T_5= -d\bar{u}$.
We summarize the three hadron flavor conventions in the last three rows of Table~\ref{tab:IUV_m_compare},
and list the pseudo-scalar octet matrices as the following,

\begin{align}
P_{\text{de Swart}}&:= 
\left(
\begin{array}{ccc}
\frac{-\pi ^0}{\sqrt{2}}-\frac{\eta_8}{\sqrt{6}} 
& -\pi ^+                                                 & K^+  \\
-\pi ^- 
& \frac{\pi ^0}{\sqrt{2}} -\frac{\eta_8}{\sqrt{6}} 
& K^0 \\
K^- 
& \bar{K}^0  
& \sqrt{\frac{2}{3}}\eta_8 \\
\end{array}
\right)\,,
\label{mat:m_compare_swart}\\
P_{\text{Chen}}&:= 
\left(
\begin{array}{ccc}
\frac{\pi ^0}{\sqrt{2}}-\frac{\eta_8}{\sqrt{6}} 
& -\pi ^+  
& K^+  \\
\pi ^-  
& -\frac{\pi ^0}{\sqrt{2}} -\frac{\eta_8}{\sqrt{6}} 
& K^0  \\
K^- 
& -\bar{K}^0 
& \sqrt{\frac{2}{3}}\eta_8 \\
\end{array}
\right)\,,\label{mat:m_compare_chen}\\
P_{\text{Rabl}}&:= 
\left(
\begin{array}{ccc}
\frac{\pi ^0}{\sqrt{2}}+\frac{\eta_8}{\sqrt{6}} 
& \pi ^+  
& K^+  \\
\pi ^- 
& -\frac{\pi ^0}{\sqrt{2}} +\frac{\eta_8}{\sqrt{6}} 
& K^0 \\
K^-
& \bar{K}^0 
& -\sqrt{\frac{2}{3}}\eta_8 \\
\end{array}
\right)\,.
\label{mat:m_compare_rabl}
\end{align}

Conventionally, the $SU_3$ octet are organized by its $SU_2$ subgroup which reflects the isospin.
As explained in Section~\eqref{sec:iso_con},
there are two possible conventions for the anti-quarks.
For the work of de Swart, the isospin doublet convention at the hadronic level is 
$\bar{K} = \left(\bar{K}^0\,\, -\!\!K^-\right)^T$,
and his flavor convention is $-K^-=T_8=-s\bar{u}$.
Both of them immediately conclude that the isospin convention at the quark level is $-\bar{u} = \ket{1/2, -1/2}$.
Thus, we reached at a quark model explanation of de Swart's hadron flavor convention.
His isospin multiplets are organized as
\begin{align}
\text{de Swart}:&\nonumber\\
    &
    \vec{\pi}:=\triplet{-\pi^+}{\pi^0}{\pi^-} = \triplet{u\bar{d}}{\frac{1}{\sqrt{2}}(-u\bar{u} + d \bar{d})}{-d\bar{u}},\\
    &K:=\doublet{K^+}{K^0} = \doublet{u\bar{s}}{d\bar{s}}, 
    \bar{K}:=\doublet{\bar{K}^0}{-K^-} = \doublet{s\bar{d}}{-s\bar{u}},
    \label{con:iso_swart}
\end{align}
where each doublet or triplet are organized by $\doublet{\ket{\frac{1}{2},\frac{1}{2}}}{\ket{\frac{1}{2},-\frac{1}{2}}}$ or $\triplet{\ket{1,1}}{\ket{1,0}}{\ket{1,-1}}$.

For comparison, we also list the isospin conventions $-\bar{d} = \ket{1/2, 1/2}$ for Chen, and Rabl.
\begin{align}
\text{Chen}:&\nonumber\\
    & 
    \vec{\pi}:=\triplet{\pi^+}{\pi^0}{\pi^-} = \triplet{-u\bar{d}}{\frac{1}{\sqrt{2}}(u\bar{u} - d \bar{d})}{d\bar{u}},\\
    & K:=\doublet{K^+}{K^0} = \doublet{u\bar{s}}{d\bar{s}},
    \bar{K}:=\doublet{\bar{K}^0}{K^-} = \doublet{-s\bar{d}}{s\bar{u}} \label{con:iso_chen_dbar}\\
\text{Rabl}:&\nonumber\\
    &
    \vec{\pi}:=\triplet{-\pi^+}{\pi^0}{\pi^-} = \triplet{-u\bar{d}}{\frac{1}{\sqrt{2}}(u\bar{u} - d \bar{d})}{d\bar{u}},\\
    &
    K:=\doublet{K^+}{K^0} = \doublet{u\bar{s}}{d\bar{s}}, 
    \bar{K}:=\doublet{-\bar{K}^0}{K^-} = \doublet{-s\bar{d}}{s\bar{u}}
    \label{con:iso_rabl_dbar}
\end{align}
Theoretically, one could also adopt the isospin conventions $-\bar{u} = \ket{1/2, -1/2}$ for Chen and Rabl. 
For completeness, we list the corresponding isospin multiplets with this convention in Appendix~\ref{sec:n_ubar}.

We need to point out that once the meson matrix (which is equivalent to adopting a hadron flavor convention) is fixed, one only needs a meson with quark component $\bar{d}$ or $\bar{u}$ in order to fix the isospin convention.
Additional assignment would either be redundant or inconsistent.
For example, the meson matrix assignment 
$P_\text{rabl}$ in Eqn.~\eqref{mat:m_compare_rabl} which is widely used in ChPT, and the convention $\ket{\pi^+} = -\ket{1,1}$ will concludes the doublet
$\bar{K} = \left(-\bar{K}^0\, K^-\right)^T$
not 
$\bar{K} = \left(\bar{K}^0\, -\!\!K^-\right)^T$.

One may argue that, despite the inconsistent assignment $\left(\bar{K}^0\, -\!\!K^-\right)^T$, a redefinition of the $\bar{K}$ field is sufficient to cease this inconsistency.
This is perhaps the reason why the convention issue does not attract enough attention.
However, not all of the parameters in the Lagrangian are free to adjust, 
especially, what appears to be the irrelevant overall phase factor in $SU_2$ is deeply connected by $SU_3$.
The sneaky redefinition can only cause confusion and misunderstanding, and we strongly suggest to do everything mathematically strict and correct.

From the Table~\ref{tab:IUV_m_compare}, one can also read horizontally and get directly the isospin multiplets.
However, if only the meson matrix is offered,
one cannot tell which descending operator conventions has been used. 
In other word, the hadron flavor convention or the meson matrix alone does not lead to the isospin convention, 
although these two conventions are closely related.
One should make a clear distinction between mathematical basis which directly furnish the representation and physical particles which one may, in principle, arbitrarily invent a convention.

Specifically, from Table~\ref{tab:IUV_m_compare},
we can see that both operator conventions $\{I_\pm, V_\pm\}$ and $\{I_\pm, U_\pm\}$
result in the same wave function for $T_6$, i.e., $T_6 = -\frac{1}{\sqrt{6}}(uu^* +dd^* -2ss^*)$.
Like the rest octet, mathematically, $T_6$ is treated as $\ket{(p,q),I,I_z,Y}= +\ket{(1,1),0,0,0}$, i.e., all the wave functions are directly identified as the $T_i$ with no further sign conventions.
These states can be marked directly by their quantum numbers in the ISFs table, like Table~II in Ref.~\cite{deSwart:1963pdg}.
To replace these quantum numbers with physical baryons and mesons, one must refer to his hadron flavor and isospin conventions.

This distinction is often not realized and the mistakes happen even in the paper which supposed to offer ISFs.
For example, the meson matrix in the paper of Rabl et.al.~\cite{Rabl:1975zy}, happens to be the same as that of the widely used in ChPT.
Since there is a non-trivial phase between $T_6$ and their $\eta_8$.
In their Table~VI, $\eta_8$ is actually supposed to the mathematical basis $T_6$ with quantum number $(I,Y)=(0,0)$, rather than physical $\eta_8$ under his convention!
Then, the sign of the ISF in the channels like $\Sigma\eta\to 10$ should be changed.
In contrast, there is no such problem when quantum numbers are used to represent for the mathematical basis,
such as what has been done in Table~II I in Ref.~\cite{Haacke:1975rt} and the tables in Ref.~\cite{deSwart:1963pdg}.

However, to do the real calculations, one has to got to the physical basis.
Once the meson matrix is fixed to be Eqn.~\eqref{mat:m_compare_rabl}, to use the tables in Rabl et.al.~\cite{Rabl:1975zy}, one has to refer to their isospin convention in Eqn.~\eqref{con:iso_rabl_dbar} or Eqn.~\eqref{con:iso_rabl_ubar},
and keeps in mind that their $\eta_8=-\ket{0,0}$.
We have also carefully checked that, the ISFs table in the Chapter 47 of review of modern physics by Particle Data Group~\cite{PDG} is a direct translation of the ISFs tables of Ref.~\cite{deSwart:1963pdg}, and the mathematical basis are rewritten into physical isospin multiplets.
As long as the isospin multiplets are explicitly defined, 
there would be no ambiguity.

The charge conjugation operator can add the additional constrain on the phases of the particle anti-particle pairs within a multiplet, concluding a meson matrix whose flavor wave function is quite symmetric.
For example, in de Swart's convention, $K^+=u\bar{s}\leftrightarrow K^-=s\bar{u}, \pi^+=-u\bar{d} \leftrightarrow \pi^- = -d\bar{u}$.
And in the convention of Rabl et.al.~\cite{Rabl:1975zy}, $K^0=d\bar{s} \leftrightarrow \bar{K^0}= s\bar{d}, \pi^+=u\bar{d} \leftrightarrow \pi^- = d\bar{u}$.
The symmetry of the wave functions will make the construction of the Lagrangian physically straightforwardly.
For instance, to construct the mass term of the mesons,
one would expect that it is proportional to $\pi^+ \pi^- + \pi^- \pi^+ + \pi^0\pi^0 +\ldots$.
However, this convenience comes at a price,
one has to keep in mind the nontrivial signs in the isospin multiplets.

In contrast, the convention from Chen et.al.~\cite{Chen:1979qz} has the advantage that the particles are directly the mathematical bases without any phase in Eqn.~\eqref{con:iso_chen_dbar}.
However, a non-trivial negative sign shows up when doing the charge conjugation.
For example, $\overline{\pi^+} =-\overline{u\bar{d}} = -d\bar{u}= -\pi^-$.
This explain the following puzzling behavior of the octet mass term,
\begin{align}
\tr(PP) &=
-2 K^0 \bar{K}^0+\left(\pi ^0\right)^2+\eta _8^2+2 K^- K^+-2 \pi ^- \pi ^+ \\
&= \pi^0 \overline{\pi^0} + \pi^- \overline{\pi^-} + \pi^+ \overline{\pi^+} + K^- \overline{K^-} + K^+ \overline{K^+} + K^0 \overline{K^0} + \bar{K}^0 \overline{\bar{K}^0} + \eta_8 \overline{\eta_8} \label{eqn:m_chen_trace}
\end{align}
In short, there is always a trade off between mathematical and physical simplicity.

\subsection{Octet Baryons}

The work flow for the octet baryons is quite different from that of mesons.
Mathematically, both baryons and anti-baryons fulfills the octet\footnote{We constrain ourselves to the octet, not the decuplet.}.
From the perspective of the $SU_N$ group theory, 
baryons and anti-baryons are the same.
But physically, we want to classify them into different multiplets because they have different baryon number.
As the start, one can directly sort the baryons and anti-baryons as what has been done in Table~\ref{tab:IUV_m_compare} by their quantum numbers $I,I_z,Y$,
\begin{align}
    & p, n, \Sigma^+, \Sigma^0, \Sigma^-, \Lambda, \Xi^0, \Xi^-\\
    & \bar{\Xi}^+, \bar{\Xi}^0, \bar{\Sigma}^+, \bar{\Sigma}^0, \bar{\Sigma}^-, \bar{\Lambda}, \bar{n}, \bar{p}
\end{align}
After that, the central topic of this paper naturally arise, what would be the consistent phase conventions?
Can we freely add signs to each of them?

The charge conjugation $\hat{C}$ will play an important role here.
For the case of mesons, $\hat{C}$ relates the meson pairs \textit{within} the octet, while in the baryon case, it relates the baryon-antibaryon pairs \textit{between} the two octets.
Thus, one can freely add signs to one octet.
This is the reason why de Swart can assign~\cite{deSwart:1963pdg} $B_3 = -\Sigma^+$ just to keep $\vec{\Sigma} \cdot \vec{\pi}$ simple,
and refuse to add the negative sign before $\Xi^-$,
which will lead to $\Xi^0K^0 - \Xi^-K^+$ (note the relative negative sign) term in the coupling to $\Lambda_0^*$.

Recall that, the widely-used $3\times 3$ matrix octet $P$ is just a compact way to represent the $8\times 1$ vector.
\begin{align}
    P = U P U^\dagger \Longleftrightarrow (P_1', \ldots, P_8')^T = M_8 (P_1, \ldots, P_8)^T,
\end{align}
where $U$ and $M_8$ are the $SU_3$ matrices in the fundamental and adjoint representation, respectively.
For the octet, we want $\hat{C}P$ to transform exactly the same as $P$.
This can be achieved by Hermitian conjugate, namely, 
complex conjugate on each elements and then take transpose of the matrix as follows,
\begin{align}
    P'^\dagger = U (P^\dagger) U^\dagger \Longleftrightarrow (P_1'^*, \ldots, P_8'^*) = (P_1^*, \ldots, P_8^*)M_8^\dagger.
\end{align}
The complex conjugate on each elements is just taking the flavor wave function into its complex conjugate.
For mesons, this is what we have been done before, such as $ud^* \to d u^*$.
For baryons, we use the physically simplest convention that the wave function of the anti-baryon is the replacement the quarks into antiquarks,
such as $p^*=\bar{p}$.

The rest to determine the mathematical basis $T_i$ under transpose.
The transpose operation seems undefined for the quarks such as $T_1 = us^*$, but this is just a shorthand notation of
\begin{align}
    us^* \equiv \left(
\begin{array}{ccc}
 0 & 0 & 1 \\
 0 & 0 & 0 \\
 0 & 0 & 0 \\
\end{array}
\right).
\end{align}
In other word, the wave functions at the quark level and the matrices are mathematically equivalent.
Then all of the basis under transformation is properly defined.

For the convention of de Swart~\cite{deSwart:1963pdg}, we have,
\begin{align}
    B_i T_i &= p T_1+ n T_2+ (-\Sigma^+) T_3 + \Sigma^0 T_4 + \Sigma^-T_5 + \Lambda T_6 + \Xi^0 T_7 + \Xi^- T_8,\\
    \hat{C}(B_i T_i) &= \bar{B}_i T_i^T = \bar{p} T_1^T+ \bar{n} T_2^T+ (-\bar{\Sigma}^-) T_3^T + \bar{\Sigma}^0 T_4^T \nonumber\\
    &\,\,\,\,\,\,+ \bar{\Sigma}^+ T_5^T + \bar{\Lambda} T_6^T + \bar{\Xi}^0 T_7^T + \bar{\Xi}^+ T_8^T,\\
    &= \bar{p} (-T_8)+ \bar{n} (T_7)+ (-\bar{\Sigma}^-) (-T_5) + \bar{\Sigma}^0 (T_4) \nonumber\\
    &\,\,\,\,\,\,+ \bar{\Sigma}^+ (-T_3) + \bar{\Lambda} T_6 + \bar{\Xi}^0 (T_2) + \bar{\Xi}^+ (-T_1)\\
    &= (-\bar{\Xi}^+) T_1 + \bar{\Xi}^0 T_2 + (-\bar{\Sigma}^+) T_3 + \bar{\Sigma}^0 T_4 + \bar{\Sigma}^-T_5 + \bar{\Lambda}T_6 + \bar{n}T_7 + (-\bar{p}) T_8.
\end{align}
This reproduces what has been claimed in the convention of his anti-baryons (cf. Equation~(17.2) in Ref.~\cite{deSwart:1963pdg}).
By performing the same calculation and noting the different transpose property of $T_i$ in $\{I_\pm, U_\pm\}$, 
we can obtain the baryon and anti-baryon matrices for other conventions.
Here we summarize these three cases as follows.

\begin{align}
\text{de Swart}:& \nonumber\\
    &B=
    \begin{pmatrix}
    	-\frac{\Sigma ^0}{\sqrt{2}}-\frac{\Lambda}{\sqrt{6}} & -\Sigma ^+  & p  \\
    	-\Sigma ^- & \frac{\Sigma ^0}{\sqrt{2}} -\frac{\Lambda}{\sqrt{6}} & n \\
    	 -\Xi^- & \Xi^0  & \sqrt{\frac{2}{3}}\Lambda \\
    \end{pmatrix},
    \bar{B} = 
    \begin{pmatrix}
    	-\frac{\bar{\Sigma} ^0}{\sqrt{2}}-\frac{\bar{\Lambda}}{\sqrt{6}} & -\bar{\Sigma} ^+  & -\bar{\Xi}^+  \\
    	-\bar{\Sigma} ^- & \frac{\bar{\Sigma} ^0}{\sqrt{2}} -\frac{\bar{\Lambda}}{\sqrt{6}} & \bar{\Xi}^0 \\
    	 \bar{p} & \bar{n}  & \sqrt{\frac{2}{3}}\bar{\Lambda} \\
    \end{pmatrix},\label{mat:B_swart}\\
\text{Chen}:& \nonumber\\
    &B=
    \begin{pmatrix}
    	\frac{\Sigma ^0}{\sqrt{2}}-\frac{\Lambda}{\sqrt{6}} & -\Sigma ^+  & p  \\
    	\Sigma ^- & -\frac{\Sigma ^0}{\sqrt{2}} -\frac{\Lambda}{\sqrt{6}} & n \\
    	 \Xi^- & -\Xi^0  & \sqrt{\frac{2}{3}}\Lambda \\
    \end{pmatrix},
    \bar{B} = 
    \begin{pmatrix}
    	\frac{\bar{\Sigma} ^0}{\sqrt{2}}-\frac{\bar{\Lambda}}{\sqrt{6}} & \bar{\Sigma} ^+  & \bar{\Xi}^+  \\
    	-\bar{\Sigma} ^- & -\frac{\bar{\Sigma} ^0}{\sqrt{2}} -\frac{\bar{\Lambda}}{\sqrt{6}} & -\bar{\Xi}^0 \\
    	 \bar{p} & \bar{n}  & \sqrt{\frac{2}{3}}\bar{\Lambda} \\
    \end{pmatrix},\label{mat:B_chen}\\ 
\text{Rabl}:& \nonumber\\
    &B= 
    \begin{pmatrix}
    	\frac{\Sigma ^0}{\sqrt{2}}+\frac{\Lambda}{\sqrt{6}} & \Sigma ^+  & p  \\
    	\Sigma ^- & -\frac{\Sigma ^0}{\sqrt{2}} +\frac{\Lambda}{\sqrt{6}} & n \\
    	 \Xi^- & \Xi^0  & -\sqrt{\frac{2}{3}}\Lambda \\
    \end{pmatrix},
    \bar{B} = 
    \begin{pmatrix}
    	\frac{\bar{\Sigma} ^0}{\sqrt{2}}+\frac{\bar{\Lambda}}{\sqrt{6}} & \bar{\Sigma} ^+  & \bar{\Xi}^+  \\
    	\bar{\Sigma} ^- & -\frac{\bar{\Sigma} ^0}{\sqrt{2}} +\frac{\bar{\Lambda}}{\sqrt{6}} & \bar{\Xi}^0 \\
    	 \bar{p} & \bar{n}  & -\sqrt{\frac{2}{3}}\bar{\Lambda} \\
    \end{pmatrix}.\label{mat:B_rabl}
\end{align}

By the construction, all of the three conventions have the property that $\tr(\bar{B}'B') = \tr(U\bar{B}U^\dagger UBU^\dagger) =\tr(\bar{B}B)$, 
which gives the mass term of the octet states.
In fact, with the octet baryon and meson matrices under each convention, one can recover some of the ISFs, for example, $\tr(BP)$ will get the right hand side of $1\to 8\otimes 8$.

Since the decuplet does not show up in the decomposition $3\otimes \bar{3}=8\oplus 1$, they cannot be organized into $3\times 3$ matrix.
Thus, their couplings to the octet baryons and mesons cannot be reproduced by taking traces of the above matrices.
However, they do show up in the decomposition, 
\begin{align}
    8\otimes 8 = 27 \oplus 10 \oplus \overline{10} \oplus 8 \oplus 8 \oplus 1,
\end{align}
this ensures that the decuplet (and the octet baryons and mesons) can be packed into $8\times 8$ matrix.
This make it possible to write of coupling such as $DBP$ (Decouplet-Baryon-Meson) by the matrix multiplication method.
We put the details of this construction in Appendix~\ref{sec:Decuplet_matrix}.

\subsection{Mixing Usage of Different Conventions}\label{sec:suggestion}

Despite the extensive usage of the ISFs by de Swart~\cite{deSwart:1963pdg}, his meson matrix as shown in Eqn.~\eqref{mat:m_compare_swart} is not widely used at present.
However, in ChPT, the meson matrix as shown in Eqn.~\eqref{mat:m_compare_rabl} is widely used.
Thus, if one use the meson matrix defined in Eqn.~\eqref{mat:m_compare_rabl} and the ISFs from de Swart~\cite{deSwart:1963pdg, PDG}, this mixing usage of different conventions could result in a misleading predictions if the isospin multiplets are not properly defined.

A common misinterpretation comes from the $\eta_8$ in the meson matrix defined in Eqn.~\eqref{mat:B_rabl}.
No matter what isospin convention one uses, 
$\bar{u}=-\ket{1/2,-1/2}$ or $\bar{d}=-\ket{1/2,1/2}$,
$T_6= -\frac{1}{\sqrt{6}}(u\bar{u}+d\bar{d}-2s\bar{s})$ should always be treat as $\ket{I,I_z,Y}= +\ket{0,0,0}$.
In both de Swart and Chen's convention, $\eta_8 = +\ket{0,0,0}$,
but for the matrix form defined in Eqn.~\eqref{mat:m_compare_rabl} widely used in ChPT,
$\eta_8 = -\ket{0,0,0}$.
Fortunately, $\eta_8$ is the singlet in the $SU_2$ group, thus, such negative sign will be no physical impact.

By comparing the meson (\ref{mat:m_compare_swart}, \ref{mat:m_compare_rabl}) and baryon matrices (\ref{mat:B_swart}, \ref{mat:B_rabl}),
we arrive at the Table~\ref{tab:mixing}.

\begin{table}[!htbp]
\centering
\setlength{\tabcolsep}{3pt}
\renewcommand{\arraystretch}{1.5}
\begin{tabular}{C|CCCCCCCC}
     & T_1 & T_2 & T_3 & T_4 & T_5 & T_6 & T_7 & T_8 \\
    \hline
    \{I_\pm, V_\pm\} & us^* & ds^* & ud^* & -\frac{1}{\sqrt{2}} (uu^* - dd^*) & -du^* & -\frac{1}{\sqrt{6}}(uu^* +dd^* -2ss^*) & sd^* & -su^*\\
    \text{Mesons} & K^+ & K^0 & \pi^+ & -\pi^0 & -\pi^- & -\eta_8 & \bar{K}^0 & -K^-\\
    \text{Baryons} & p & n & \Sigma^+ & -\Sigma^0 & -\Sigma^- & -\Lambda & \Xi^0 & -\Xi^-\\
    \hline
\end{tabular}
    \caption{The relation of physical particles and mathematical basis when mixing the usage of meson \eqref{mat:m_compare_rabl} and baryon\eqref{mat:B_rabl} matrix with the isoscalar factors in de Swart convention\cite{deSwart:1963pdg, PDG}.}
    \label{tab:mixing}
\end{table}

From Table~\ref{tab:mixing}, we can see that the isospin multiplets has to be defined as the following,
\begin{align}
    K&:=\doublet{K^+}{K^0}, 
    \vec{\pi}:=\triplet{\pi^+}{-\pi^0}{-\pi^-},
    \bar{K}:=\doublet{\bar{K}^0}{-K^-}, 
    -\eta_8:= \ket{0,0} \\
    N&:=\doublet{p}{n}, 
    \vec{\Sigma}:=\triplet{\Sigma^+}{-\Sigma^0}{-\Sigma^-},
    \Xi:=\doublet{\Xi^0}{-\Xi^-}, 
    -\Lambda:= \ket{0,0}
    \label{con:mixing}
\end{align}

As we have stated before, the baryon and meson matrix is nothing but a convenient way to organizing the octets.
In principle, one is not bothered to explicitly write down the meson and baryon matrix, if the correct ISFs and isospin multiplets are used, as what has been done by de Swart.

\section{Isoscalar Factors}\label{sec:isf}

The isoscalar factors are the agents between small group $SU_2$ and a larger group $SU_3$.
With ISFs and the Clebsch-Gordan coefficients (CGCs) of the smaller group at hand, the CGCs of the bigger group can be constructed.
In some sense, ISFs are not as fundamental as CGCs, 
since the physical processes are directly linked with CGCs, which physicists directly work with.
In the case of $SU_3$, the ISFs only appear when one intends to separate the contributions of the $SU_2$ isospin group but still wants to find the relations between the couplings of different $SU_3$ flavor multiplets, 
much like how the famous Wigner-Eckart theorem helps us separate the dynamics from the geometry.

Here we demonstrate the process of getting $SU_3$ CGCs from the quark level with the Young/Weyl tableaux method, 
where the anti-quarks are represented by the anti-symmetrized combination of quarks.
The phase conventions follow that of Ref.~\cite{Chen:2002gd}, where the detailed calculation steps can be found.

\Yvcentermath0
\Yboxdim{10pt}
\begin{table}[!htbp]
\centering
\setlength{\tabcolsep}{1pt}
\renewcommand{\arraystretch}{2}
\begin{tabular}{LL|CCCCCC}
        &   & (\young(uu,d),\young(ud,s) ) & (\young(uu,d),\young(us,d) ) & (\young(uu,s),\young(ud,d) ) & (\young(ud,d),\young(uu,s) ) & (\young(ud,s),\young(uu,d) ) & (\young(us,d),\young(uu,d) ) \\
        &   & (p, \pi^0)        & (p, \eta_8)       & (\Sigma^+, K^0)   & (n,\pi^+)         & (\Sigma^0, K^+)   & (\Lambda, K^+)    \\
        &   & (K^+, \Sigma^0)   & (K^+, \Lambda)    & (\pi^+, n)        & (K^0,\Sigma^+)    & (\pi^0, p)        & (\eta_8, p)        \\
    \hline 
                &   \young(uuud,ds)    & \frac{1}{\sqrt{3}}     & 0                     & \frac{1}{\sqrt{6}}   & \frac{1}{\sqrt{6}}    & \frac{1}{\sqrt{3}}        & 0                       \\
                &   \young(uuus,dd)    & -\frac{1}{2 \sqrt{15}} & \frac{3}{2 \sqrt{5}}  & \frac{1}{\sqrt{30}}  & \frac{1}{\sqrt{30}}   & -\frac{1}{2 \sqrt{15}}    & \frac{3}{2 \sqrt{5}}    \\ 
    \Delta^+    &   \young(uuud,d,s)   & \frac{1}{\sqrt{3}}     & 0                     & -\frac{1}{\sqrt{6}}  & \frac{1}{\sqrt{6}}    & -\frac{1}{\sqrt{3}}       & 0                       \\ 
                &   \young(uuu,dds)    & \frac{1}{2 \sqrt{3}}   & \frac{1}{2}           & \frac{1}{\sqrt{6}}   & -\frac{1}{\sqrt{6}}   & -\frac{1}{2 \sqrt{3}}     & -\frac{1}{2}            \\
    p           &   \young(uuu,dd,s)   & \sqrt{\frac{3}{14}}    & -\frac{1}{\sqrt{14}}  & 0                    & -\sqrt{\frac{3}{7}}   & 0                         & \sqrt{\frac{2}{7}}      \\
    p           &   \young(uuu,dd,s)   & \sqrt{\frac{2}{105}}   & 2 \sqrt{\frac{2}{35}} & -\sqrt{\frac{7}{15}} & -\frac{2}{\sqrt{105}} & \sqrt{\frac{7}{30}}       & -\frac{1}{\sqrt{70}}    
\end{tabular}
\caption{One part of $SU_3$ CGCs}
\label{tab:P8CGC}
\end{table}
\Yvcentermath1
\Yboxdim{12pt}

In Table~\ref{tab:P8CGC}, we list two possible interpretations of the Weyl tableaux in the decay particles, namely baryon first or meson first convention.
The two conventions originate from the fact that both octet baryons and mesons live in the same $SU_3$ representation.
For example, the Weyl tableaux $\young(uu,d)$ which stands for one state in octet can be identified with $K^+$ or proton.
This two-fold role of the Weyl tableaux turns out to be very useful.
In order to get the table, we also identify, say, the $\young(uu,d)$ with $\young(uuu,dd,s)$, where a $SU_3$ flavor vacuum $\young(u,d,s)$ is prepended to the tableaux.
The ISFs in Ref.~\cite{PDG} adopt the baryon first convention in the above table, such as, $p \to p \pi^0$.

We interpret the term in the Lagrangian like $\bar{B}_1 B_2 M_3$ to be directly related to the $B_2 +M_3 \to B_1$,
whose Hermitian conjugate reflects the "decay" process $B_1 \to B_2 + M_3$.

As has been explained in the end of Section~\ref{sec:su3cgc}, in order to distinguish the two protons (which are in the last two rows of Table ~\ref{tab:P8CGC}), the two possible couplings $8 \to 8 \otimes 8$ can be further classified into symmetric $8_1$ and anti-symmetric $8_2$ part.
The symmetrizer (anti-symmetrizer) can be assigned to $B\leftrightarrow B$ or $B \leftrightarrow M$.
Because of the following property,
\begin{align}
\tr(\{\bar{B}, B\} M) = \tr(\bar{B}, \{B, M\}),\,\,\, \tr([\bar{B}, B] M) = \tr(\bar{B}, [B, M]).
\end{align}
In the language of the Young tableaux, the above is a special case of the following,
\begin{align}
    \yng(2,1)\otimes\yng(2,1)&= \yng(4,2)\oplus\yng(4,1,1)\oplus\yng(3,3)\oplus\yng(3,2,1)\oplus\yng(3,2,1)\oplus\yng(2,2,2)\,,\\
8 \otimes 8 &= 27 \oplus 10 \oplus \overline{10} \oplus 8 \oplus 8 \oplus 1.
\end{align}

Unfortunately, the $SU_3$ CGCs in the Table~\ref{tab:P8CGC} does not fulfill this requirement.
For example, in the coupled channel of proton, when exchanging $\young(uu,d)\ \young(ud,s) \leftrightarrow \young(ud,s)\ \young(uu,d)$ or equivalently, $p\pi^0 \leftrightarrow \Sigma^0 K^+$, 
the CGCs change like $\sqrt{3/14} \leftrightarrow 0$ or $\sqrt{2/105} \leftrightarrow \sqrt{7/30}$, which is neither symmetric nor anti-symmetric.
However, additional rotation between the last two row vector will solve this issue,
\begin{align}
  \doublet{p_1}{p_2} &\to \left[\matrixTwo{\cos{\theta}}{-\sin{\theta}}{\sin{\theta}}{\cos{\theta}} \matrixTwo{\sqrt{3/14}}{0}{\sqrt{2/105}}{\sqrt{7/30}}\right]\doublet{p\pi^0}{\Sigma^0 k^+}\\
    &= \matrixTwo{x}{x}{y}{-y} \doublet{p\pi^0}{\Sigma^0 k^+}
\end{align}
where $x,y$ are constants to be determined later.
Note that one cannot fix those constants ahead, such as $x=y=1/\sqrt{2}$, 
since $p\pi^0, \Sigma^0 k^+$ are not the only channels which proton can couple to.

There are two solutions for the equation, 
$\theta_1 = \pi -\arctan{(3/\sqrt{5})}, \theta_2 =-\arctan{(3/\sqrt{5})}$, which will lead to,  
\begin{align}
\theta_1 = \pi -\arctan{\frac{3}{\sqrt{5}}} &\Rightarrow 
  \doublet{p_1}{p_2} \to \matrixTwo{-\sqrt{\frac{3}{20}}}{-\sqrt{\frac{3}{20}}}{\sqrt{\frac{1}{12}}}{-\sqrt{\frac{1}{12}}} \doublet{p\pi^0}{\Sigma^0 k^+},\\
\theta_2 = -\arctan{\frac{3}{\sqrt{5}}} &\Rightarrow 
  \doublet{p_1'}{p_2'} \to \matrixTwo{\sqrt{\frac{3}{20}}}{\sqrt{\frac{3}{20}}}{-\sqrt{\frac{1}{12}}}{\sqrt{\frac{1}{12}}} \doublet{p\pi^0}{\Sigma^0 k^+}.
\end{align}
The two solutions of $\theta$ only differ by an overall negative sign.
Since we use the order convention that symmetric combination is before that of the anti-symmetric one,
the first non-zero coefficient of the symmetric combination should positive, 
which leads to the second rotation angle.
Note that this rotation angle is universal for all $8\to 8\otimes 8$ couplings.

\begin{table}[!htbp]
\centering
{\renewcommand\arraystretch{2}
	\begin{tabular}{C|CCCC}
        & (N, \pi)        & (N, \eta_8)       & (\Sigma, K)    & (\Lambda, K)    \\
    \hline
    \Delta    & \frac{1}{\sqrt{2}} & 0 & -\frac{1}{\sqrt{2}} & 0 \\
    N_1 & \frac{3}{2 \sqrt{5}} & \frac{1}{2 \sqrt{5}} & -\frac{3}{2 \sqrt{5}} & \frac{1}{2 \sqrt{5}} \\
    N_2 & -\frac{1}{2} & \frac{1}{2} & -\frac{1}{2} & -\frac{1}{2} \\
	\end{tabular}
 }
\caption{The isoscalar factor after symmetrization and anti-symmetrization of $8\otimes 8$.
}
\label{tab:ISF_temp}
\end{table}

To get the $SU_3$ ISF, we need to divide the $SU_3$ CGCs with the corresponding $SU_2$ isospin CGCs, which results in Table~\ref{tab:ISF_temp}.
In Table~\ref{tab:ISF_temp}, we replaced the particles in Table~\ref{tab:P8CGC} with their isospin families,
and dropped the rows of the Weyl tableaux beyond the octet and decuplet baryons, such as $\young(uuud,ds)$.
Please note that one isospin channel in Table~\ref{tab:ISF_temp} corresponds to several charged channels in Table~\ref{tab:P8CGC}, for example, $\Delta^+\to p\pi^0$ and $\Delta^+ \to n\pi^+$ belong to the family $\Delta \to N \pi$.

Strictly speaking, the isospin multiplets such as $\vec{\pi}$ need to be defined.
However, in Chen's convention, the isospin multiplets are directly identified by the Weyl tableaux without additional phases, i.e., $\vec{\pi} =\left(\pi^+\,\, \pi^0\,\, \pi^-\right)^T$ which directly corresponds to the isospin states, $\left(+\ket{1,1}\,+\ket{1,0}\,+\ket{1,-1}\right)^T$.

In the Table~\ref{tab:P8CGC}, the quark components are fixed to be $u,u,u,d,d,s$.
Nothing stops us to explore other quark components, like $u,u,u,u,d,s$, and calculating the corresponding ISFs.
Following along this line, we list the ISFs in Chen's convention in the following.

\begin{align}
1 \to 8 \otimes 8\nonumber\\
\left(\begin{array}{c}{\Lambda^*_0}\end{array}\right) &\rightarrow \left(\begin{array}{cccc} {N \bar{K}} & {\Sigma \pi} & {\Lambda \eta} & {\Xi K}\end{array} \right) = \frac{1}{\sqrt{8}}\left(\begin{array}{cccc} 2 & -3 & 1 & -2 \end{array}\right)^{1/2} \label{tab:isf_188}\,,\\
8_1 \to 8 \otimes 8\nonumber\\
\left(\begin{array}{c}{N} \\ {\Sigma} \\ {\Lambda} \\ {\Xi}\end{array}\right)
 &\xrightarrow{D}\left(\begin{array}{rrrrr}{N \pi} & {N \eta} & {\Sigma K} & {\Lambda K} \\ {N \bar{K}} & {\Sigma \pi} & {\Lambda \pi} & {\Sigma \eta} & {\Xi K} \\ {N \bar{K}} & {\Sigma \pi} & {\Lambda \eta} & {\Xi K} \\ {\Sigma \bar{K}} & {\Lambda \bar{K}} & {\Xi \pi} & {\Xi \eta}\end{array}\right)
=\frac{1}{\sqrt{20}}\left(\begin{array}{rrrrr} {9} & {1} & {-9} & {1} & \\ {6} & {0} & {-4} & {-4} & {6} \\ {2} & {12} & {4} & {-2} & \\ {9} & {1} & {-9} & {1} & \end{array}\right)^{1 / 2}\,,\label{tab:isf_8188}\\
8_2 \to 8 \otimes 8\nonumber\\
\left(\begin{array}{c}{N} \\ {\Sigma} \\ {\Lambda} \\ {\Xi}\end{array}\right)
 &\xrightarrow{F}\left(\begin{array}{rrrrr}{N \pi} & {N \eta} & {\Sigma K} & {\Lambda K} \\ {N \bar{K}} & {\Sigma \pi} & {\Lambda \pi} & {\Sigma \eta} & {\Xi K} \\ {N \bar{K}} & {\Sigma \pi} & {\Lambda \eta} & {\Xi K} \\ {\Sigma \bar{K}} & {\Lambda \bar{K}} & {\Xi \pi} & {\Xi \eta}\end{array}\right)
=\frac{1}{\sqrt{12}}\left(\begin{array}{rrrrr}{-3} & {3} & {-3} & {-3} &  \\ {2} & {-8} & {0} & {0} & {-2} \\ {-6} & {0} & {0} & {-6} \\ {-3} & {3} & {-3} & {-3}\end{array}\right)^{1 / 2}\,,\label{tab:isf_8288}\\
10 \to 8 \otimes 8\nonumber\\
\left(\begin{array}{c}{\Delta} \\ {\Sigma^*} \\ {\Xi^*} \\ {\Omega}\end{array}\right) 
&\rightarrow\left(\begin{array}{rrrrr}{N \pi} & {\Sigma K} & {} \\ {N \bar{K}} & {\Sigma \pi} & {\Lambda \pi} & {\Sigma\eta} & {\Xi K}\\ {\Sigma \bar{K}} & {\Lambda \bar{K}} & {\Xi \pi} & {\Xi \eta} \\ {\Xi\bar{K}} & {} \end{array}\right)
=\frac{1}{\sqrt{12}}\left(\begin{array}{rrrrr} {6} & {-6} \\ {2}  & {2} & {3} & {-3} & {-2} \\ {3} & {3} & {3} & {-3} \\ {12} & {}\end{array}\right)^{1 / 2}\,,\label{tab:isf_1088}\\
8 \to 10 \otimes 8\nonumber\\
\left(\begin{array}{c}{N} \\ {\Sigma} \\ {\Lambda} \\ {\Xi}\end{array}\right)
&\rightarrow\left(\begin{array}{rrrr}{\Delta \pi} &  {\Sigma^* K} & {} \\ {\Delta \bar{K}} & {\Sigma^* \pi} & {\Sigma^* \eta} & {\Xi K} \\ {\Sigma^*\pi} & {\Xi^* K}\\{\Sigma^* \bar{K}} & {\Xi^* \pi} & {\Xi^* \eta} & {\Omega K}\end{array}\right) 
=\frac{1}{\sqrt{15}}\left(\begin{array}{rrrr} {12} &{3}  \\ {8} & {-2} & {3} & {-2} \\ {9} & {6} \\ {3} & {-3} & {3} & {-6}\end{array}\right)^{1 / 2}\,,\label{tab:isf_8108}\\
10 \to 10 \otimes 8\nonumber\\
\left(\begin{array}{c}{\Delta} \\ {\Sigma^*} \\ {\Xi^*} \\ {\Omega}\end{array}\right) 
&\rightarrow\left(\begin{array}{rrrr}{\Delta \pi} & {\Delta \eta} & {\Sigma^* K} \\ {\Delta \bar{K}} & {\Sigma^* \pi} & {\Sigma^* \eta} & {\Xi^* K} \\ {\Sigma^* \bar{K}} & {\Xi^* \pi} & {\Xi^* \eta} & {\Omega K} \\ {\Xi^* \bar{K}} & {\Omega \eta}\end{array}\right) 
=\frac{1}{\sqrt{24}}\left(\begin{array}{rrrr} {15} & {-3} & {6} \\ {8} & {8} & {0} & {8} \\ {12} & {3} & {3} & {6} \\ {12} & {12}\end{array}\right)^{1 / 2}\,, \label{tab:isf_10108}
\end{align}

As a specific example to exhibit the effect of choosing different conventions, 
in Table~\ref{tab:isf_1088},
we see that the ISFs of the highest decuplet 
$\Delta \to p\pi$ and $\Delta \to \Sigma K$ are different.
In Chen's convention, 
the ISFs of $\Delta \to p\pi$ should be positive, since $p$ is higher than $\Sigma$.
However, with Haacke and Rabl's convention, 
$I(\Sigma)=1$ which is larger than $I(p) =1/2$, 
so the ISF of $\Delta \to \Sigma K$ should be positive.
Although we agree on the same set of descending operators $\{I_-, U_-\}$, 
we have a distinct convention on the highest weight.
This difference will cause an overall negative sign to the CGCs (or equivalently, ISFs) on $10 \to 8\otimes 8$, 
as it should be,
since the relative signs within the multiplets are controlled by the same set of descending operators $\{I_-, U_-\}$.

If both the descending operator set and the highest state conventions are different, then apart from the overall phase differences, the ISFs of each $SU_2$ multiplets within each $SU_3$ multiplets could also be different.
Since all conventions should be mathematically equivalent,
this superficially contradicting results can be absorbed by the redefinition of the $SU_2$ isospin multiplets.

Specifically, there is a similar ISFs table in PDG~\cite{PDG} whose absolute value is the same as what we got but the signs are different.
To reproduce it, 
we can redefine the fields of $N,K,\Lambda, \eta_8$, and change the overall sign of $1\to 8\otimes 8, 8_2 \to 8\otimes 8$.

The above discussion also offers a way to check the consistency of different conventions.
If the ISFs among different conventions are still different after redefinition of the whole $SU_2$ multiplets and the whole $SU_3$ coupled channels,
then at least one convention is not self-consistent.
Note that this consistency checker is a necessary not sufficient condition.

There is a subtlety when translating the ISFs to the $A\to B\otimes C$ form when $B$ and $C$ are both in octet.
Since mathematically, octet baryons and mesons share the same quantum numbers in group irreps, one has to assign a convention to distinguish them.
We hereby adopt the baryon-first-convention, namely, 
\begin{align}
    \left(\begin{array}{cc|c}
\stackrel{8}{I_1 Y_1} & \stackrel{8}{I_2 Y_2} & \stackrel{\mu_\gamma}{I Y}
\end{array}\right). 
\end{align}
It is in interpreted as $\mathrm{Baryon}(I,Y) \to \mathrm{Baryon}(I_1,Y_1) \otimes \mathrm{Meson} (I_2,Y_2)$ instead of $\mathrm{Baryon} \to \mathrm{Meson} (I_1,Y_1) \otimes \mathrm{Baryon}(I_2,Y_2)$.

This order convention is important when building the Lagrangian from the ISFs and $SU_2$ CGCs, especially when the Lagrangian is written in the charged states.
For example, with the baryon-first convention, the $pp\pi^0$ Lagrangian should be proportional to $\bra{p\pi^0}\ket{p}=\bra{\frac{1}{2},\frac{1}{2};1,0}\ket{\frac{1}{2},\frac{1}{2}}=\frac{1}{\sqrt{3}}$ instead of 
$\bra{1,0;\frac{1}{2},\frac{1}{2}}\ket{\frac{1}{2},\frac{1}{2}}=-\frac{1}{\sqrt{3}}$.
Due to this order convention, in theory, one has to be cautious when adopting coupling constants from various sources.
In practice, however, this subtlety is often unnoticeable.
Since the couplings are conventionally reorganized into isospin multiplets, where $SU_2$ CGCs (thus the order convention) are implicitly included which eliminates the order ambiguity.
For instance, $\Delta\Delta \pi$ vertex is often expressed as
$\bar{\Delta}_\mu \gamma^5\gamma^\nu \vec{T}\Delta^\mu \partial_\nu \vec{\pi}$
where each vector component of $\vec{T}$ is a $4\times 4$ matrix with $SU_2$ CGCs included~\cite{Ronchen:2012eg}.

\section{Summary}\label{sec:summary}

In this paper, we tracked and compared possible conventions in the construction of the Lagrangian at hadronic level.
We pointed out that these conventions can be classified into two different sources.
One source is from the group theory, where people may choose different ways to generalize the $SU_2$ Cordon-Shortley phase convention to $SU_3$.
We also give a group theory explanation that the Baird-Biedenharn convention is more natural than the widely used de Swart convention.
The second sources of the conventions are pure notational, and they arise at the identification stage, such as whether the isospin of $\pi^+$ should be identified as $\ket{1,1}$ or $-\ket{1,1}$.

By detailed analysis of three different conventions,
we pointed out some common misconceptions about the sign convention of $\eta_8$ and also give some suggestions when one wants to mix the results from different conventions.

The tools to track the conventions are the quark model which servers as the agent to translate the abstract mathematical basis into physical visions.
It also has the ability to check various conventions at finer details, and we suggest to use it to check the consistency of all conventions.

\section{Acknowledgment}

Yu Lu is grateful to Professor Jia-Lun Ping, Dr. Yu-Fei Wang, and Dr. Mao-Jun Yan for helpful discussions. This work is supported by the National Natural Science Foundation of China under Grant Nos. 12175239 and 12221005,
and by the National Key Research and Development Program of China under Contracts 2020YFA0406400,
and by the Chinese Academy of Sciences under Grant No. YSBR-101, 
and by the Xiaomi Foundation / Xiaomi Young Talents Program.

\section{Appendix}\label{appendix}

\subsection{Octet Wave Functions Under de Swart Convention}\label{sec:wf_swart}

Here we list the matrix elements and the steps to get he $SU_3$ flavor wave functions of the octet under de Swart convention \cite{deSwart:1963pdg},
i.e., the matrix elements of $I_\pm, V_\pm$ are positive.
These wave functions are not the octet meson wave functions, because of the additional hadron flavor conventions.

The corresponding $I_\pm, V_\pm, U_\pm$ matrix elements in $(p,q) = (1,0)$ and $(1,1)$ representation are
\begin{align}
&(p,q) = (1,0):\nonumber \\
&I_- = \begin{pmatrix}
 0 & 0 & 0 \\
 1 & 0 & 0 \\
 0 & 0 & 0 \\
\end{pmatrix},
V_- = 
\begin{pmatrix}
 0 & 0 & 0 \\
 0 & 0 & 0 \\
 1 & 0 & 0 \\
\end{pmatrix},
U_- = 
\begin{pmatrix}
 0 & 0 & 0 \\
 0 & 0 & 0 \\
 0 & 1 & 0 \\
\end{pmatrix}
\\
&(p,q) = (1,1):\nonumber \\
I_- &= 
\begin{pmatrix}
 0 & 0 & 0 & 0 & 0 & 0 & 0 & 0 \\
 1 & 0 & 0 & 0 & 0 & 0 & 0 & 0 \\
 0 & 0 & 0 & 0 & 0 & 0 & 0 & 0 \\
 0 & 0 & \sqrt{2} & 0 & 0 & 0 & 0 & 0 \\
 0 & 0 & 0 & \sqrt{2} & 0 & 0 & 0 & 0 \\
 0 & 0 & 0 & 0 & 0 & 0 & 0 & 0 \\
 0 & 0 & 0 & 0 & 0 & 0 & 0 & 0 \\
 0 & 0 & 0 & 0 & 0 & 0 & 1 & 0 \\
\end{pmatrix},
V_- = \left(
\begin{array}{cccccccc}
 0 & 0 & 0 & 0 & 0 & 0 & 0 & 0 \\
 0 & 0 & 0 & 0 & 0 & 0 & 0 & 0 \\
 0 & 0 & 0 & 0 & 0 & 0 & 0 & 0 \\
 \frac{1}{\sqrt{2}} & 0 & 0 & 0 & 0 & 0 & 0 & 0 \\
 0 & 1 & 0 & 0 & 0 & 0 & 0 & 0 \\
 \sqrt{\frac{3}{2}} & 0 & 0 & 0 & 0 & 0 & 0 & 0 \\
 0 & 0 & 1 & 0 & 0 & 0 & 0 & 0 \\
 0 & 0 & 0 & \frac{1}{\sqrt{2}} & 0 & \sqrt{\frac{3}{2}} & 0 & 0 \\
\end{array}
\right) \nonumber\\
U_- &= 
\begin{pmatrix}
 0 & 0 & 0 & 0 & 0 & 0 & 0 & 0 \\
 0 & 0 & 0 & 0 & 0 & 0 & 0 & 0 \\
 -1 & 0 & 0 & 0 & 0 & 0 & 0 & 0 \\
 0 & -\frac{1}{\sqrt{2}} & 0 & 0 & 0 & 0 & 0 & 0 \\
 0 & 0 & 0 & 0 & 0 & 0 & 0 & 0 \\
 0 & \sqrt{\frac{3}{2}} & 0 & 0 & 0 & 0 & 0 & 0 \\
 0 & 0 & 0 & \frac{1}{\sqrt{2}} & 0 & -\sqrt{\frac{3}{2}} & 0 & 0 \\
 0 & 0 & 0 & 0 & 1 & 0 & 0 & 0 \\
\end{pmatrix}
\end{align}
The ascending operators can be get by taking transpose of the descending operators, namely, $I_+ = I_-^T, V_+ = V_-^T, U_+= U_-^T$.
With the matrix form of the descending operators, we can enumerate the octet and get their flavor wave functions as follows,
\begin{align}
    \ket{8^{[1]}} := \ket{us^*}& \\
\hat{I}_- \ket{8^{[1]}} &= \hat{I}_ - \ket{us^*}\\
\Rightarrow \ket{8^{[2]}} &= \ket{ds^*}\\
    \hat{U}_- \ket{8^{[1]}} &= \hat{U}_-\ket{us^*}\\
    - \ket{8^{[3]}} &= \ket{(\hat{U}_-u)s^*} + \ket{u (-\hat{U}_+s)^*}\\
    &= 0 + \ket{u (-d)^*}\\
\Rightarrow \ket{8^{[3]}} &= \ket{u d^*}\\
    \hat{I}_- \ket{8^{[3]}} &= \hat{I}_ - \ket{ud^*}\\
    \sqrt{2} \ket{8^{4}} &= \ket{dd^*}  + \ket{u (-I_+d)^*}\\
    &= \ket{dd^*}  + \ket{u u^*}\\
\Rightarrow \ket{8^{[4]}} &= -\frac{1}{\sqrt{2}} (\ket{uu^*} - \ket{dd^*})\\
    \hat{I}_- \ket{8^{[4]}} &= -\frac{1}{\sqrt{2}} [I_-(\ket{uu^*} - \ket{dd^*})]\\
    \sqrt{2} \ket{8^{[5]}} &= -\frac{1}{\sqrt{2}} (\ket{du^*}+\ket{du^*})\\
\Rightarrow \ket{8^{[5]}} &= - \ket{du^*}\\
    \hat{V}_- \ket{8^{[1]}} &= \hat{V}_- \ket{us^*}\\
    \frac{1}{\sqrt{2}} \ket{8^{[4]}} + \sqrt{\frac{3}{2}} \ket{8^{[6]}}&= \ket{ss^*} - \ket{uu^*}\\
\Rightarrow \ket{8^{[6]}} &= \frac{1}{\sqrt{6}} \ket{-uu^* -dd^* +2 ss^*}\\
    \hat{V}_- \ket{8^{[3]}} &= \hat{V}_- \ket{ud^*}\\
\Rightarrow \ket{8^{[7]}} &= \ket{sd^*} \\
    \hat{I}_- \ket{8^{[7]}} &= \hat{I}_-\ket{sd^*}\\
\Rightarrow \ket{8^{[8]}} &= - \ket{su^*}
\end{align}

\subsection{Isospin Convention on $\bar{u}$}\label{sec:n_ubar}

The isospin multiplets under the isospin convention $\bar{u} := -\ket{\frac{1}{2},-\frac{1}{2}}$ for Chen et.al.~\cite{Chen:1979qz} and Rabl et.al.~\cite{Rabl:1975zy}.
\begin{align}
\text{Chen}&:\nonumber\\
    &
    \vec{\pi}:=\triplet{-\pi^+}{-\pi^0}{-\pi^-} = \triplet{u\bar{d}}{-\frac{1}{\sqrt{2}}(u\bar{u} - d \bar{d})}{-d\bar{u}},\nonumber\\
    &K:=\doublet{K^+}{K^0} = \doublet{u\bar{s}}{d\bar{s}}, 
    \bar{K}:=\doublet{-\bar{K}^0}{-K^-} = \doublet{s\bar{d}}{-s\bar{u}}\label{con:iso_chen_ubar}\\
\text{Rabl}&:\nonumber\\
    &
    \vec{\pi}:=\triplet{\pi^+}{-\pi^0}{-\pi^-} = \triplet{u\bar{d}}{-\frac{1}{\sqrt{2}}(u\bar{u} - d \bar{d})}{-d\bar{u}},\nonumber\\
    &K:=\doublet{K^+}{K^0} = \doublet{u\bar{s}}{d\bar{s}}, 
    \bar{K}:=\doublet{\bar{K}^0}{-K^-} = \doublet{s\bar{d}}{-s\bar{u}} \label{con:iso_rabl_ubar}
\end{align}

\subsection{Matrix Form of the Decuplet}\label{sec:Decuplet_matrix}

As an example, we provide a matrix form in this appendix that includes the coupling of baryon decuplet. 
To achieve this goal, one first needs to introduce the following ten matrices $D^\alpha~(\alpha=1,\cdots10)$, under the convention of Swart:
\arraycolsep=1.5pt
\begin{align}
        &D^1=\left(
\begin{array}{cccccccc}
 0 & 0 & 0 & 0 & -\frac{1}{\sqrt{2}} & 0 & 0 & 0 \\
 0 & 0 & 0 & 0 & 0 & 0 & 0 & 0 \\
 0 & 0 & 0 & 0 & 0 & 0 & 0 & \frac{1}{\sqrt{2}} \\
 0 & 0 & 0 & 0 & 0 & 0 & 0 & 0 \\
 0 & 0 & 0 & 0 & 0 & 0 & 0 & 0 \\
 0 & 0 & 0 & 0 & 0 & 0 & 0 & 0 \\
 0 & 0 & 0 & 0 & 0 & 0 & 0 & 0 \\
 0 & 0 & 0 & 0 & 0 & 0 & 0 & 0 \\
\end{array}
\right),\quad
D^2=\left(
\begin{array}{cccccccc}
 0 & 0 & 0 & \frac{1}{\sqrt{3}} & 0 & 0 & 0 & 0 \\
 0 & 0 & 0 & 0 & -\frac{1}{\sqrt{6}} & 0 & 0 & 0 \\
 0 & 0 & 0 & 0 & 0 & 0 & -\frac{1}{\sqrt{6}} & 0 \\
 0 & 0 & 0 & 0 & 0 & 0 & 0 & \frac{1}{\sqrt{3}} \\
 0 & 0 & 0 & 0 & 0 & 0 & 0 & 0 \\
 0 & 0 & 0 & 0 & 0 & 0 & 0 & 0 \\
 0 & 0 & 0 & 0 & 0 & 0 & 0 & 0 \\
 0 & 0 & 0 & 0 & 0 & 0 & 0 & 0 \\
\end{array}
\right),\nonumber\\
&D^3=\left(
\begin{array}{cccccccc}
 0 & 0 & -\frac{1}{\sqrt{6}} & 0 & 0 & 0 & 0 & 0 \\
 0 & 0 & 0 & \frac{1}{\sqrt{3}} & 0 & 0 & 0 & 0 \\
 0 & 0 & 0 & 0 & 0 & 0 & 0 & 0 \\
 0 & 0 & 0 & 0 & 0 & 0 & -\frac{1}{\sqrt{3}} & 0 \\
 0 & 0 & 0 & 0 & 0 & 0 & 0 & \frac{1}{\sqrt{6}} \\
 0 & 0 & 0 & 0 & 0 & 0 & 0 & 0 \\
 0 & 0 & 0 & 0 & 0 & 0 & 0 & 0 \\
 0 & 0 & 0 & 0 & 0 & 0 & 0 & 0 \\
\end{array}
\right),\quad
D^4=\left(
\begin{array}{cccccccc}
 0 & 0 & 0 & 0 & 0 & 0 & 0 & 0 \\
 0 & 0 & -\frac{1}{\sqrt{2}} & 0 & 0 & 0 & 0 & 0 \\
 0 & 0 & 0 & 0 & 0 & 0 & 0 & 0 \\
 0 & 0 & 0 & 0 & 0 & 0 & 0 & 0 \\
 0 & 0 & 0 & 0 & 0 & 0 & -\frac{1}{\sqrt{2}} & 0 \\
 0 & 0 & 0 & 0 & 0 & 0 & 0 & 0 \\
 0 & 0 & 0 & 0 & 0 & 0 & 0 & 0 \\
 0 & 0 & 0 & 0 & 0 & 0 & 0 & 0 \\
\end{array}
\right),\nonumber\\
&D^5=\left(
\begin{array}{cccccccc}
 0 & \frac{1}{\sqrt{6}} & 0 & 0 & 0 & 0 & 0 & 0 \\
 0 & 0 & 0 & 0 & 0 & 0 & 0 & 0 \\
 0 & 0 & 0 & -\frac{1}{2 \sqrt{3}} & 0 & -\frac{1}{2} & 0 & 0 \\
 0 & 0 & 0 & 0 & -\frac{1}{2 \sqrt{3}} & 0 & 0 & 0 \\
 0 & 0 & 0 & 0 & 0 & 0 & 0 & 0 \\
 0 & 0 & 0 & 0 & -\frac{1}{2} & 0 & 0 & 0 \\
 0 & 0 & 0 & 0 & 0 & 0 & 0 & \frac{1}{\sqrt{6}} \\
 0 & 0 & 0 & 0 & 0 & 0 & 0 & 0 \\
\end{array}
\right),\quad
D^6=\left(
\begin{array}{cccccccc}
 -\frac{1}{2 \sqrt{3}} & 0 & 0 & 0 & 0 & 0 & 0 & 0 \\
 0 & \frac{1}{2 \sqrt{3}} & 0 & 0 & 0 & 0 & 0 & 0 \\
 0 & 0 & \frac{1}{2 \sqrt{3}} & 0 & 0 & 0 & 0 & 0 \\
 0 & 0 & 0 & 0 & 0 & -\frac{1}{2} & 0 & 0 \\
 0 & 0 & 0 & 0 & -\frac{1}{2 \sqrt{3}} & 0 & 0 & 0 \\
 0 & 0 & 0 & \frac{1}{2} & 0 & 0 & 0 & 0 \\
 0 & 0 & 0 & 0 & 0 & 0 & -\frac{1}{2 \sqrt{3}} & 0 \\
 0 & 0 & 0 & 0 & 0 & 0 & 0 & \frac{1}{2 \sqrt{3}} \\
\end{array}
\right),\nonumber\\
&D^7=\left(
\begin{array}{cccccccc}
 0 & 0 & 0 & 0 & 0 & 0 & 0 & 0 \\
 -\frac{1}{\sqrt{6}} & 0 & 0 & 0 & 0 & 0 & 0 & 0 \\
 0 & 0 & 0 & 0 & 0 & 0 & 0 & 0 \\
 0 & 0 & \frac{1}{2 \sqrt{3}} & 0 & 0 & 0 & 0 & 0 \\
 0 & 0 & 0 & \frac{1}{2 \sqrt{3}} & 0 & -\frac{1}{2} & 0 & 0 \\
 0 & 0 & -\frac{1}{2} & 0 & 0 & 0 & 0 & 0 \\
 0 & 0 & 0 & 0 & 0 & 0 & 0 & 0 \\
 0 & 0 & 0 & 0 & 0 & 0 & -\frac{1}{\sqrt{6}} & 0 \\
\end{array}
\right),\quad
D^8=\left(
\begin{array}{cccccccc}
 0 & 0 & 0 & 0 & 0 & 0 & 0 & 0 \\
 0 & 0 & 0 & 0 & 0 & 0 & 0 & 0 \\
 \frac{1}{\sqrt{6}} & 0 & 0 & 0 & 0 & 0 & 0 & 0 \\
 0 & \frac{1}{2 \sqrt{3}} & 0 & 0 & 0 & 0 & 0 & 0 \\
 0 & 0 & 0 & 0 & 0 & 0 & 0 & 0 \\
 0 & \frac{1}{2} & 0 & 0 & 0 & 0 & 0 & 0 \\
 0 & 0 & 0 & -\frac{1}{2 \sqrt{3}} & 0 & -\frac{1}{2} & 0 & 0 \\
 0 & 0 & 0 & 0 & -\frac{1}{\sqrt{6}} & 0 & 0 & 0 \\
\end{array}
\right),\nonumber\\
&D^9=\left(
\begin{array}{cccccccc}
 0 & 0 & 0 & 0 & 0 & 0 & 0 & 0 \\
 0 & 0 & 0 & 0 & 0 & 0 & 0 & 0 \\
 0 & 0 & 0 & 0 & 0 & 0 & 0 & 0 \\
 \frac{1}{2 \sqrt{3}} & 0 & 0 & 0 & 0 & 0 & 0 & 0 \\
 0 & \frac{1}{\sqrt{6}} & 0 & 0 & 0 & 0 & 0 & 0 \\
 -\frac{1}{2} & 0 & 0 & 0 & 0 & 0 & 0 & 0 \\
 0 & 0 & \frac{1}{\sqrt{6}} & 0 & 0 & 0 & 0 & 0 \\
 0 & 0 & 0 & \frac{1}{2 \sqrt{3}} & 0 & -\frac{1}{2} & 0 & 0 \\
\end{array}
\right),\quad
D^{10}=\left(
\begin{array}{cccccccc}
 0 & 0 & 0 & 0 & 0 & 0 & 0 & 0 \\
 0 & 0 & 0 & 0 & 0 & 0 & 0 & 0 \\
 0 & 0 & 0 & 0 & 0 & 0 & 0 & 0 \\
 0 & 0 & 0 & 0 & 0 & 0 & 0 & 0 \\
 0 & 0 & 0 & 0 & 0 & 0 & 0 & 0 \\
 0 & 0 & 0 & 0 & 0 & 0 & 0 & 0 \\
 \frac{1}{\sqrt{2}} & 0 & 0 & 0 & 0 & 0 & 0 & 0 \\
 0 & \frac{1}{\sqrt{2}} & 0 & 0 & 0 & 0 & 0 & 0 \\
\end{array}
\right).
\end{align}

Then, we can organize the baryon decuplet matrix $D\equiv D^\alpha \mathcal{B}_\alpha$ with
\begin{align}
\mathcal{B}_\alpha=\{\Delta^{++},\Delta^{+},\Delta^{0},\Delta^-,\Sigma^{*+},\Sigma^{*0},\Sigma^{*-},\Xi^{0},\Xi^{-},\Omega\}_\alpha.
\end{align}

Similarly, one can construct an anti-baryon decuplet matrix $\bar{D}\equiv \bar{D}_\alpha \bar{\mathcal{B}}^\alpha$ with
\begin{align}
    \bar{D}_\alpha=\left(D^\alpha\right)^T,\quad \bar{\mathcal{B}}^\alpha=\{\bar{\Delta}^{--},\bar{\Delta}^{-},\bar{\Delta}^{0},\bar{\Delta}^+,\bar{\Sigma}^{*-},\bar{\Sigma}^{*0},\bar{\Sigma}^{*+},\bar{\Xi}^{0},\bar{\Xi}^+,\bar{\Omega}\}^\alpha.
\end{align}


In addition, in order to construct the octet meson and baryon matrices, we also need to introduce the following two types of matrices, denoted as $O_A^{a}$ and $O_S^{a}~(a=1,\cdots,8)$:
\begin{align}
&O_A^1=\left(
\begin{array}{cccccccc}
 0 & 0 & 0 & -\frac{1}{2 \sqrt{3}} & 0 & -\frac{1}{2} & 0 & 0 \\
 0 & 0 & 0 & 0 & -\frac{1}{\sqrt{6}} & 0 & 0 & 0 \\
 0 & 0 & 0 & 0 & 0 & 0 & -\frac{1}{\sqrt{6}} & 0 \\
 0 & 0 & 0 & 0 & 0 & 0 & 0 & -\frac{1}{2 \sqrt{3}} \\
 0 & 0 & 0 & 0 & 0 & 0 & 0 & 0 \\
 0 & 0 & 0 & 0 & 0 & 0 & 0 & -\frac{1}{2} \\
 0 & 0 & 0 & 0 & 0 & 0 & 0 & 0 \\
 0 & 0 & 0 & 0 & 0 & 0 & 0 & 0 \\
\end{array}
\right),\quad O_A^2=\left(
\begin{array}{cccccccc}
 0 & 0 & \frac{1}{\sqrt{6}} & 0 & 0 & 0 & 0 & 0 \\
 0 & 0 & 0 & \frac{1}{2 \sqrt{3}} & 0 & -\frac{1}{2} & 0 & 0 \\
 0 & 0 & 0 & 0 & 0 & 0 & 0 & 0 \\
 0 & 0 & 0 & 0 & 0 & 0 & -\frac{1}{2 \sqrt{3}} & 0 \\
 0 & 0 & 0 & 0 & 0 & 0 & 0 & -\frac{1}{\sqrt{6}} \\
 0 & 0 & 0 & 0 & 0 & 0 & \frac{1}{2} & 0 \\
 0 & 0 & 0 & 0 & 0 & 0 & 0 & 0 \\
 0 & 0 & 0 & 0 & 0 & 0 & 0 & 0 \\
\end{array}
\right),\nonumber\\
&O_A^3=\left(
\begin{array}{cccccccc}
 0 & -\frac{1}{\sqrt{6}} & 0 & 0 & 0 & 0 & 0 & 0 \\
 0 & 0 & 0 & 0 & 0 & 0 & 0 & 0 \\
 0 & 0 & 0 & -\frac{1}{\sqrt{3}} & 0 & 0 & 0 & 0 \\
 0 & 0 & 0 & 0 & -\frac{1}{\sqrt{3}} & 0 & 0 & 0 \\
 0 & 0 & 0 & 0 & 0 & 0 & 0 & 0 \\
 0 & 0 & 0 & 0 & 0 & 0 & 0 & 0 \\
 0 & 0 & 0 & 0 & 0 & 0 & 0 & -\frac{1}{\sqrt{6}} \\
 0 & 0 & 0 & 0 & 0 & 0 & 0 & 0 \\
\end{array}
\right),\quad O_A^4=\left(
\begin{array}{cccccccc}
 \frac{1}{2 \sqrt{3}} & 0 & 0 & 0 & 0 & 0 & 0 & 0 \\
 0 & -\frac{1}{2 \sqrt{3}} & 0 & 0 & 0 & 0 & 0 & 0 \\
 0 & 0 & \frac{1}{\sqrt{3}} & 0 & 0 & 0 & 0 & 0 \\
 0 & 0 & 0 & 0 & 0 & 0 & 0 & 0 \\
 0 & 0 & 0 & 0 & -\frac{1}{\sqrt{3}} & 0 & 0 & 0 \\
 0 & 0 & 0 & 0 & 0 & 0 & 0 & 0 \\
 0 & 0 & 0 & 0 & 0 & 0 & \frac{1}{2 \sqrt{3}} & 0 \\
 0 & 0 & 0 & 0 & 0 & 0 & 0 & -\frac{1}{2 \sqrt{3}} \\
\end{array}
\right),\nonumber\\
&O_A^5=\left(
\begin{array}{cccccccc}
 0 & 0 & 0 & 0 & 0 & 0 & 0 & 0 \\
 \frac{1}{\sqrt{6}} & 0 & 0 & 0 & 0 & 0 & 0 & 0 \\
 0 & 0 & 0 & 0 & 0 & 0 & 0 & 0 \\
 0 & 0 & \frac{1}{\sqrt{3}} & 0 & 0 & 0 & 0 & 0 \\
 0 & 0 & 0 & \frac{1}{\sqrt{3}} & 0 & 0 & 0 & 0 \\
 0 & 0 & 0 & 0 & 0 & 0 & 0 & 0 \\
 0 & 0 & 0 & 0 & 0 & 0 & 0 & 0 \\
 0 & 0 & 0 & 0 & 0 & 0 & \frac{1}{\sqrt{6}} & 0 \\
\end{array}
\right),\quad O_A^6=\left(
\begin{array}{cccccccc}
 \frac{1}{2} & 0 & 0 & 0 & 0 & 0 & 0 & 0 \\
 0 & \frac{1}{2} & 0 & 0 & 0 & 0 & 0 & 0 \\
 0 & 0 & 0 & 0 & 0 & 0 & 0 & 0 \\
 0 & 0 & 0 & 0 & 0 & 0 & 0 & 0 \\
 0 & 0 & 0 & 0 & 0 & 0 & 0 & 0 \\
 0 & 0 & 0 & 0 & 0 & 0 & 0 & 0 \\
 0 & 0 & 0 & 0 & 0 & 0 & -\frac{1}{2} & 0 \\
 0 & 0 & 0 & 0 & 0 & 0 & 0 & -\frac{1}{2} \\
\end{array}
\right),\nonumber\\
&O_A^7=\left(
\begin{array}{cccccccc}
 0 & 0 & 0 & 0 & 0 & 0 & 0 & 0 \\
 0 & 0 & 0 & 0 & 0 & 0 & 0 & 0 \\
 \frac{1}{\sqrt{6}} & 0 & 0 & 0 & 0 & 0 & 0 & 0 \\
 0 & \frac{1}{2 \sqrt{3}} & 0 & 0 & 0 & 0 & 0 & 0 \\
 0 & 0 & 0 & 0 & 0 & 0 & 0 & 0 \\
 0 & -\frac{1}{2} & 0 & 0 & 0 & 0 & 0 & 0 \\
 0 & 0 & 0 & -\frac{1}{2 \sqrt{3}} & 0 & \frac{1}{2} & 0 & 0 \\
 0 & 0 & 0 & 0 & -\frac{1}{\sqrt{6}} & 0 & 0 & 0 \\
\end{array}
\right),\quad O_A^8=\left(
\begin{array}{cccccccc}
 0 & 0 & 0 & 0 & 0 & 0 & 0 & 0 \\
 0 & 0 & 0 & 0 & 0 & 0 & 0 & 0 \\
 0 & 0 & 0 & 0 & 0 & 0 & 0 & 0 \\
 \frac{1}{2 \sqrt{3}} & 0 & 0 & 0 & 0 & 0 & 0 & 0 \\
 0 & \frac{1}{\sqrt{6}} & 0 & 0 & 0 & 0 & 0 & 0 \\
 \frac{1}{2} & 0 & 0 & 0 & 0 & 0 & 0 & 0 \\
 0 & 0 & \frac{1}{\sqrt{6}} & 0 & 0 & 0 & 0 & 0 \\
 0 & 0 & 0 & \frac{1}{2 \sqrt{3}} & 0 & \frac{1}{2} & 0 & 0 \\
\end{array}
\right),
\end{align}
\begin{align}
&O_S^1=\left(
\begin{array}{cccccccc}
 0 & 0 & 0 & -\frac{\sqrt{\frac{3}{5}}}{2} & 0 & \frac{1}{2 \sqrt{5}} & 0 & 0 \\
 0 & 0 & 0 & 0 & -\sqrt{\frac{3}{10}} & 0 & 0 & 0 \\
 0 & 0 & 0 & 0 & 0 & 0 & \sqrt{\frac{3}{10}} & 0 \\
 0 & 0 & 0 & 0 & 0 & 0 & 0 & \frac{\sqrt{\frac{3}{5}}}{2} \\
 0 & 0 & 0 & 0 & 0 & 0 & 0 & 0 \\
 0 & 0 & 0 & 0 & 0 & 0 & 0 & -\frac{1}{2 \sqrt{5}} \\
 0 & 0 & 0 & 0 & 0 & 0 & 0 & 0 \\
 0 & 0 & 0 & 0 & 0 & 0 & 0 & 0 \\
\end{array}
\right),\quad O_S^2=\left(
\begin{array}{cccccccc}
 0 & 0 & \sqrt{\frac{3}{10}} & 0 & 0 & 0 & 0 & 0 \\
 0 & 0 & 0 & \frac{\sqrt{\frac{3}{5}}}{2} & 0 & \frac{1}{2 \sqrt{5}} & 0 & 0 \\
 0 & 0 & 0 & 0 & 0 & 0 & 0 & 0 \\
 0 & 0 & 0 & 0 & 0 & 0 & \frac{\sqrt{\frac{3}{5}}}{2} & 0 \\
 0 & 0 & 0 & 0 & 0 & 0 & 0 & \sqrt{\frac{3}{10}} \\
 0 & 0 & 0 & 0 & 0 & 0 & \frac{1}{2 \sqrt{5}} & 0 \\
 0 & 0 & 0 & 0 & 0 & 0 & 0 & 0 \\
 0 & 0 & 0 & 0 & 0 & 0 & 0 & 0 \\
\end{array}
\right),\nonumber\\
&O_S^3=\left(
\begin{array}{cccccccc}
 0 & \sqrt{\frac{3}{10}} & 0 & 0 & 0 & 0 & 0 & 0 \\
 0 & 0 & 0 & 0 & 0 & 0 & 0 & 0 \\
 0 & 0 & 0 & 0 & 0 & -\frac{1}{\sqrt{5}} & 0 & 0 \\
 0 & 0 & 0 & 0 & 0 & 0 & 0 & 0 \\
 0 & 0 & 0 & 0 & 0 & 0 & 0 & 0 \\
 0 & 0 & 0 & 0 & \frac{1}{\sqrt{5}} & 0 & 0 & 0 \\
 0 & 0 & 0 & 0 & 0 & 0 & 0 & -\sqrt{\frac{3}{10}} \\
 0 & 0 & 0 & 0 & 0 & 0 & 0 & 0 \\
\end{array}
\right),\quad O_S^4=\left(
\begin{array}{cccccccc}
 -\frac{\sqrt{\frac{3}{5}}}{2} & 0 & 0 & 0 & 0 & 0 & 0 & 0 \\
 0 & \frac{\sqrt{\frac{3}{5}}}{2} & 0 & 0 & 0 & 0 & 0 & 0 \\
 0 & 0 & 0 & 0 & 0 & 0 & 0 & 0 \\
 0 & 0 & 0 & 0 & 0 & -\frac{1}{\sqrt{5}} & 0 & 0 \\
 0 & 0 & 0 & 0 & 0 & 0 & 0 & 0 \\
 0 & 0 & 0 & -\frac{1}{\sqrt{5}} & 0 & 0 & 0 & 0 \\
 0 & 0 & 0 & 0 & 0 & 0 & \frac{\sqrt{\frac{3}{5}}}{2} & 0 \\
 0 & 0 & 0 & 0 & 0 & 0 & 0 & -\frac{\sqrt{\frac{3}{5}}}{2} \\
\end{array}
\right),\nonumber\\
&O_S^5=\left(
\begin{array}{cccccccc}
 0 & 0 & 0 & 0 & 0 & 0 & 0 & 0 \\
 -\sqrt{\frac{3}{10}} & 0 & 0 & 0 & 0 & 0 & 0 & 0 \\
 0 & 0 & 0 & 0 & 0 & 0 & 0 & 0 \\
 0 & 0 & 0 & 0 & 0 & 0 & 0 & 0 \\
 0 & 0 & 0 & 0 & 0 & -\frac{1}{\sqrt{5}} & 0 & 0 \\
 0 & 0 & \frac{1}{\sqrt{5}} & 0 & 0 & 0 & 0 & 0 \\
 0 & 0 & 0 & 0 & 0 & 0 & 0 & 0 \\
 0 & 0 & 0 & 0 & 0 & 0 & \sqrt{\frac{3}{10}} & 0 \\
\end{array}
\right),\quad O_S^6=\left(
\begin{array}{cccccccc}
 \frac{1}{2 \sqrt{5}} & 0 & 0 & 0 & 0 & 0 & 0 & 0 \\
 0 & \frac{1}{2 \sqrt{5}} & 0 & 0 & 0 & 0 & 0 & 0 \\
 0 & 0 & -\frac{1}{\sqrt{5}} & 0 & 0 & 0 & 0 & 0 \\
 0 & 0 & 0 & -\frac{1}{\sqrt{5}} & 0 & 0 & 0 & 0 \\
 0 & 0 & 0 & 0 & -\frac{1}{\sqrt{5}} & 0 & 0 & 0 \\
 0 & 0 & 0 & 0 & 0 & \frac{1}{\sqrt{5}} & 0 & 0 \\
 0 & 0 & 0 & 0 & 0 & 0 & \frac{1}{2 \sqrt{5}} & 0 \\
 0 & 0 & 0 & 0 & 0 & 0 & 0 & \frac{1}{2 \sqrt{5}} \\
\end{array}
\right),\nonumber\\
&O_S^7=\left(
\begin{array}{cccccccc}
 0 & 0 & 0 & 0 & 0 & 0 & 0 & 0 \\
 0 & 0 & 0 & 0 & 0 & 0 & 0 & 0 \\
 \sqrt{\frac{3}{10}} & 0 & 0 & 0 & 0 & 0 & 0 & 0 \\
 0 & \frac{\sqrt{\frac{3}{5}}}{2} & 0 & 0 & 0 & 0 & 0 & 0 \\
 0 & 0 & 0 & 0 & 0 & 0 & 0 & 0 \\
 0 & \frac{1}{2 \sqrt{5}} & 0 & 0 & 0 & 0 & 0 & 0 \\
 0 & 0 & 0 & \frac{\sqrt{\frac{3}{5}}}{2} & 0 & \frac{1}{2 \sqrt{5}} & 0 & 0 \\
 0 & 0 & 0 & 0 & \sqrt{\frac{3}{10}} & 0 & 0 & 0 \\
\end{array}
\right),\quad O_S^8=\left(
\begin{array}{cccccccc}
 0 & 0 & 0 & 0 & 0 & 0 & 0 & 0 \\
 0 & 0 & 0 & 0 & 0 & 0 & 0 & 0 \\
 0 & 0 & 0 & 0 & 0 & 0 & 0 & 0 \\
 \frac{\sqrt{\frac{3}{5}}}{2} & 0 & 0 & 0 & 0 & 0 & 0 & 0 \\
 0 & \sqrt{\frac{3}{10}} & 0 & 0 & 0 & 0 & 0 & 0 \\
 -\frac{1}{2 \sqrt{5}} & 0 & 0 & 0 & 0 & 0 & 0 & 0 \\
 0 & 0 & -\sqrt{\frac{3}{10}} & 0 & 0 & 0 & 0 & 0 \\
 0 & 0 & 0 & -\frac{\sqrt{\frac{3}{5}}}{2} & 0 & \frac{1}{2 \sqrt{5}} & 0 & 0 \\
\end{array}
\right).
\end{align}

By employing two types of matrices, $O_A^a$ and $O_S^a$, one can construct meson and baryon octet matrices as follows:
\begin{equation}
    \begin{aligned}
    &\Phi_{A/S}\equiv O_{A/S}^a \phi_a,\quad\, \phi_a=\{K^+,K^0,-\pi^+,\pi^0,\pi^-,\eta_8,\bar{K}^0,-K^-\}_a,\\
    &B_{A/S}\equiv O_{A/S}^a B_a,\quad B_a=\{p,n,-\Sigma^+,\Sigma^0,\Sigma^-,\Lambda,\Xi^0,\Xi^-\}_a,\\
    &\bar{B}_{A/S}\equiv O_{A/S}^a \bar{B}_a,\quad \bar{B}_a=\{-\bar{\Xi}^-,\bar{\Xi}^0,-\bar{\Sigma}^+,\bar{\Sigma}^0,\bar{\Sigma}^-,\bar{\Lambda},\bar{n},-\bar{p}\}_a.
    \end{aligned}
\end{equation}

Based on the introduced matrices above, we can construct the interaction vertices involving the meson octet, baryon octet, and baryon decuplet in a unified form. 
At the leading order, there are the following seven independent structures:
\begin{align}
    \text{Mass term :}\quad & \mathcal{L}_{1} \propto\left\langle\Phi_A \Phi_A\right\rangle=\left\langle\Phi_S \Phi_S\right\rangle\text{~or~}\left\langle\bar{B}_A B_A\right\rangle=\left\langle\bar{B}_S B_S\right\rangle\quad\text{(Meson/Baryon octet)} \\
& \mathcal{L}_{2} \propto\langle\bar{D}D\rangle,\\
\text{Yukawa term :}\quad & \mathcal{L}_3 \propto\left\langle\bar{B}_A B_A \Phi_A\right\rangle=\left\langle\bar{B}_A B_S \Phi_S\right\rangle=\left\langle\bar{B}_S B_A \Phi_S\right\rangle=\left\langle\bar{B}_S B_S \Phi_A\right\rangle . \\
& \mathcal{L}_4 \propto\left\langle\bar{B}_S B_S \Phi_S\right\rangle=-\frac{5}{3}\left\langle\bar{B}_A B_A \Phi_S\right\rangle=-\frac{5}{3}\left\langle\bar{B}_A B_S \Phi_A\right\rangle=-\frac{5}{3}\left\langle\bar{B}_S B_A \Phi_A\right\rangle,\\
\text{Decuplet term :}\quad & \mathcal{L}_5 \propto\left\langle\bar{D} B_S \Phi_S\right\rangle=-\frac{2}{\sqrt{5}}\left\langle\bar{D} B_A \Phi_S\right\rangle=\frac{2}{\sqrt{5}}\left\langle\bar{D} B_S \Phi_A\right\rangle \\
& \mathcal{L}_6 \propto\left\langle\bar{B}_S D \Phi_S\right\rangle=-\frac{2}{\sqrt{5}}\left\langle\bar{B}_A D \Phi_S\right\rangle=\frac{2}{\sqrt{5}}\left\langle\bar{B}_S D \Phi_A\right\rangle \\
& \mathcal{L}_7 \propto\left\langle\bar{D} D\Phi_S\right\rangle=\frac{1}{\sqrt{5}}\left\langle\bar{D} D\Phi_A\right\rangle,
\end{align}
where the brackets $\langle\dots\rangle$ represent taking the trace of the matrix. Interaction vertices not mentioned above, such as $\left\langle\bar{D} B_A \Phi_A\right\rangle$, are all zero. 
It can be verified that the interaction vertices obtained from the above Lagrangian are consistent with those derived from the SU${}_3$ CGCs.

\bibliographystyle{utphys}
\bibliography{refs}

\end{document}